\documentclass[aps,prd,12pt,eqsecnum,showpacs,preprintnumbers,nofootinbib,superscriptaddress]{revtex4}
\usepackage{amssymb,amsmath,amsthm,graphicx,amscd,color,ulem,enumerate}

\begin{document}

\title{Thermodynamics of quantum systems strongly coupled to a heat bath I. Operator thermodynamic functions and relations}


\author{J.-T. Hsiang}
\email{cosmology@gmail.com}
\affiliation{Center for Field Theory and Particle Physics, Department of Physics,\\
Fudan University, Shanghai 200433, China}
\author{B. L. Hu}
\email{blhu@umd.edu}
\affiliation{Maryland Center for Fundamental Physics and Joint Quantum Institute,
University of Maryland, College Park, Maryland 20742-4111, USA}

\date{\today}


\begin{abstract}
The thermodynamics of small quantum many-body systems strongly coupled to a heat bath at low temperatures with non-Markovian behavior are new challenges for quantum thermodynamics, as traditional thermodynamics is built on large systems vanishingly weakly coupled to a non-dynamical reservoir. Important also are the quantum attributes,  as in quantum coherence, correlations,  entanglement and fluctuations. All told, one needs to reexamine the meaning of the thermodynamic functions, the viability of the thermodynamic relations and the validity of the thermodynamic laws anew. In one popular approach to quantum thermodynamics the closed system,  comprising the system  of interest and the bath it is strongly coupled to, is assumed to be in a global thermal state throughout. In this set-up three theories of thermodynamics at strong coupling have been proposed, those of Gelin \& Thoss \cite{GT09}, Seifert \cite{Se16} and Jarzynski \cite{JA17}.  This paper provides a quantum formulation of them, with Jarzynski's two different representations encompassing the former two. Operator thermodynamic potentials and thermodynamic relations are presented. We mention issues related to energy and entropy in the two representations, a possible way to define quantum work in our functional formulation, and how to connect with the open system nonequilibrium dynamics approach to quantum thermodynamics, as proposed in \cite{HC17}.   
 
\end{abstract}
\maketitle

\tableofcontents

\section{Introduction}

\subsection{New challenges in quantum thermodynamics} 

Small quantum many-body systems strongly coupled to a heat reservoir at low temperatures are the new focuses of interest for quantum thermodynamics \cite{Mahler}. Under these hitherto lesser explored conditions, one needs to re-examine the meaning of the thermodynamic functions, the viability of the thermodynamic relations and the validity of the thermodynamic laws anew. Traditional thermodynamics is built on large systems weakly coupled to a reservoir \cite{SpoLeb}, and for quantum systems, only the spin-statistics aspect is studied, as in quantum statistical mechanics, leaving the important factors of quantum coherence, correlations, entanglement and fluctuations as new challenges for quantum thermodynamics. To see the difference strong coupling makes, the  definition of  heat, as the energy transferred between the system and the reservoir, for systems strongly coupled to a bath, is nontrivial\footnote{Only for systems in a steady state is it well defined, when there is no net change in the coupling energy, thus the attributions of the coupling energy to the system or to the reservoir or any partiality, are all equivalent.}.  Esposito et al \cite{EspOchGal}, for example,  show that any heat definition expressed as an energy change in the reservoir energy plus any fraction of the system-reservoir interaction is not an exact differential when evaluated along reversible isothermal transformations, except when that fraction is zero. Even in that latter case the reversible heat divided by temperature, namely entropy, does not satisfy the third law of thermodynamics and diverges in the low temperature limit. For quantum systems, as pointed out by Ankerhold and Pekola \cite{AnkPek}, in actual measurements, especially for solid state structures, quantum correlations between system and reservoir may be of relevance not only far from but also close to and in thermal equilibrium. Even in the weak coupling regime, this heat flow is substantial at low temperatures and may become comparable to typical predictions for the work based on conventional weak coupling approaches. It further depends sensitively on the non-Markovian features of the reservoir.  These observations exemplify  the intricacies involved in defining heat for strongly-coupled systems and added complexities for quantum systems, especially at low temperatures.

This incertitude regarding heat translates to ambiguity in the definition of thermodynamic functions  and  the thermodynamic relations. {For example, it was shown \cite{NA02,HI06,HI08,IH09,PEP13,HA11} that the expressions for the specific heat derived from the internal energies of a quantum-mechanical harmonic oscillator bilinearly coupled to a harmonic bath calculated by two different approaches  can have dramatically different behavior in the low temperature regime}. To illustrate this point,  Gelin and Thoss \cite{GT09} compared these two approaches {of calculating the internal energy of the system}, which give identical results if the system-bath coupling is negligible, but predict  significantly differently for finite system-bath coupling. In the first approach, the mean energy of the system given by the expectation value of the system Hamiltonian is evaluated with respect to the total (system+bath) canonical equilibrium distribution. The second approach is based on the partition function of the system, $Z_s$, which is postulated to be given as the ratio of the total (system+bath) and the bath partition functions. Gelin and Thoss \cite{GT09} introduce a bath-induced interaction operator $\hat{\Delta}_s$, which would account for the effects of finite system-bath coupling and analyze the two approaches for several different systems including several quantum and classical point particles and nonlinear system bath coupling. They found that Approach I\!I leads to very different results from Approach I, their differences exist already within classical mechanics, provided the system-bath interaction is not bilinear and/or the system of interest consists of more than a single particle.

{Similar ambiguity appears in the entropy of the quantum system in the same setup. In the first approach, the von Neumann entropy is chosen to be the entropy of the system, while in the second approach, the entropy is given by the derivative of the system's free energy with respect to the inverse temperature. Both definitions are equivalent in the limit of weak system-bath coupling. It has been noticed that \cite{NA02,HI06,OC06,HB08} the von Neumann entropy may not vanish when the bath temperature is close to zero, even for a simple quantum system that consists of harmonic oscillators. This nonvanishing behavior of the von Neumann entropy is related to the quantum entanglement between the system and the bath~\cite{JB04,HL09,ELvdB,DeffLutz}. On the other hand,  the entropy defined in the second approach of the same system gives an expected vanishing result, consistent with the physical picture described by~\cite{HZ95}.}

\subsection{Two major approaches and set-ups} 

{For classical thermodynamics at strong system-bath coupling, the above mentioned approaches have been systematically developed and extended by Seifert~\cite{Se16} and Jarzynski \cite{JA17} covering  more thermodynamic quantities than the internal energy in the formulation of the laws of thermodynamics. However, corresponding quantum-mechanical formulations of these two approaches would be highly desirable.}
To provide a better orientation of where our work fits in, we mention two major approaches and set-ups undertaken in current studies of strong-coupling thermodynamics for quantum systems~\cite{GT09,NA02,HI08,PEP13,OC06,HB08,Mahler,PA16,Deutsch91,Srednicki94,GLTZ06,PSW06,LPSW09,SF12,Reimann08,PSSV11,CR10,GE16,Carrega}:

1) Approach based on the assumption that the combined system $\mathbf{S}$ + bath $\mathbf{B}$, which we call \textit{the composite $\mathbf{C}$,  is in a globally thermal state} (CGTs): The composite evolves unitarily. Assumed in many recent work this set-up has the advantage that 
\begin{enumerate}[a)]
	\item it is easier to transcribe the laws of ordinary thermodynamics, such as with the use of partition function \footnote{Note, however, the pitfalls, one such  pointed out by Esposito et. al. \cite{PEP13} : If one proceeds from assuming that the composite \textbf{C} (the combined system + environment) is in a thermal state,  the behavior of the heat capacity of the system is different when it is derived from the energy of the central system at equilibrium or from a partition function approach \cite{HI06, HI08}.} 
	\item mechanical work can be unambiguously defined (see e.g., Seifert \cite{Se16}).  
\end{enumerate}
However, as pointed out by~\cite{SFTH,HC17}, in the CGTs set up, even though the closed system is assumed to be in a global thermal state, the system is not necessarily in a thermal state. One needs extra assumptions, such as the system is very weakly coupled to the bath, which undermines the purpose of strong coupling thermodynamics investigations.

2) Open system nonequilibrium dynamics (ONEq) approach: One begins by allowing a system in some initial state to interact with its environment,  follows its dynamics to late times, then examines if a steady state exists (equilibration), or further explores if the system thermalizes.  These conditions depend on the structures of the system, the properties of the bath and the way they interact. These factors need to be considered  before one can begin to construct thermodynamical quantities, build the thermodynamical relations and examine whether the well-established thermodynamical laws in traditional (weak-coupling) thermodynamics remain valid for interacting quantum many-body systems. 
\begin{enumerate}[a)]
 \item The open-system nonequilibrium dynamics approach makes no de facto reference to the partition function. 

 \item One sees how the environment exerts its influence on the system as it evolves in time.  There are well established methods like the influence functional formalism whereby one can identify the noise in the environment, derive the (stochastic) equations of motion for the open system dynamics and study the environmental effects on the system, such as dissipation, decoherence and (dis)entanglement effects. 
\end{enumerate}
 
3) It will be very useful to \textit{establish connections between the two approaches} delineated above, since each has its special advantages which could illuminate different aspects of the new challenges posed by strong coupling thermodynamics. From prior work based on the quantum Brownian model \cite{SFTH,HC17} we know the following: For strong coupling between the system and bath, if the system can approach the equilibrium state, then the reduced density matrix of the open systems is the same as the reduced density matrix in the CGTs framework upon integrating out the bath\footnote{Note, however, the final global states are different, despite the fact in both cases the dynamics is generated by the same Hamiltonian $H_c= H_s + H_i + H_b$, e.g. \eqref{E:nrjtbrs}. This is because two global systems start out with different initial states, and the unitary evolution does not change the distinguishability between the states.}. 

As far as quantum thermodynamics is concerned it would be very useful to find out if there exists any \textit{thermodynamic function which remains valid under nonequilibrium conditions}. This is almost impossible to hope for, since there is a big divide between nonequilibrium and equilibrium systems.  The key observation  focuses on whether a nonequilibrium quantum system relaxes to an equilibrium state, and if so, what physical quantities remains well defined from the nonequilibrium to the equilibrium states.

\subsection{Goal and Findings of present work} The goal of this paper is to generalize some representative classical formulations of strong coupling thermodynamics in the CGTs set-up to quantum systems under the same conditions. For this we have succeeded in finding the operator thermodynamic function and relations in a quantum reformulation of  Gelin \& Thoss (G\&T)~\cite{GT09} and a quantum formulation of Seifert~\cite{Se16}, and of Jarzynski~\cite{JA17}. Note Jarzynski's formulation for classical systems includes that of Seifert, and of G\&T.

This paper is organized as follows: In Sec.~\ref{S:rtbgfjd} we give a quick summary of the familiar thermodynamic relations, which we call traditional or weak coupling (wc) thermodynamics (TD), if only to establish notations. We then consider interacting quantum systems with the help of the Hamiltonian of mean force \cite{HTL11}. We present two formulations of thermodynamic functions and relations, one by G\&T \cite{GT09}, which has a quantum formulation, and the other by Seifert \cite{Se16} for classical systems. In Sec.~\ref{S:jrtbsgsgd} we present a quantum formulation of Seifert's thermodynamics. We mention outstanding issues in the properties of energy and entropy in these two formulations. In Sec.~\ref{S:tbwjrjc} we give a brief description of Jarzynski's \cite{JA17}  thermodynamics at strong coupling for classical systems, as a primer for our quantum formulation. In Sec.~\ref{S:tubfgdgfd} we present a quantum formulation of Jarzynski's thermodynamics in the `bare' and `partial molar' representations, which correspond to G\&T and Seifert's thermodynamics, respectively. We conclude in Sec.~\ref{S:tihbgd} with some suggestion to further developments of these theories, and how to build connections between the closed system in a global thermal state approach explored here (backed by a huge literature, see Jarzynski \cite{JA17} and references therein ) and  the open-system nonequilibrium dynamics approach to quantum thermodynamics proposed recently (see \cite{HC17} and references therein). Readers who are familiar with the classical formulations of G\&T, Seifert and Jarzynski can skip over to Sec.~\ref{S:jrtbsgsgd} and~\ref{S:tubfgdgfd}, where quantum formulations are presented.

\section{Thermodynamic Functions,  Hamiltonian of Mean Force}\label{S:rtbgfjd}

We first summarize the familiar traditional thermodynamic relations, if only to establish notations. We then consider interacting quantum systems with the help of the Hamiltonian of mean force\footnote{Hamiltonian of mean force is a useful yet not indispensable concept for this purpose. It is handy because in the same representation, the formal expressions associated with it resembles the counterparts in the traditional weak coupling thermodynamics)}. We present two formulations of thermodynamic functions and relations, one by Gelin and Thoss \cite{GT09}, which has a quantum formulation and the other by Seifert \cite{Se16} for classical systems. As we will see, they fall under the two representations of Jarzynski \cite{JA17} who formulated thermodynamics at strong coupling also  for classical systems. This we will present in the next section. With the abundance of thermodynamic quantities a word about notations is helpful: quantum expectation values or classical ensemble averages are denoted by math calligraphic, quantum operators associated with the variable $O$ will carry an overhat $\hat{O}$.

\subsection{Traditional (weak-coupling) thermodynamic relations} 

The pre-conditions for the traditional weak-coupling (wc) thermodynamic (TD) theory to be well-defined and operative for a classical or quantum system are very specific despite its wide ranging applicability: a) A system \textbf{S} of relatively few degrees of freedom is in contact with a thermal bath of a large number or infinite degrees of freedom\footnote{We shall consider only heat but no particle transfer here and thus the TD refers only to canonical, not grand canonical ensembles}; b) the coupling between the system and the bath is vanishingly small, and c) the system is eternally in a thermal equilibrium state by proxy with the bath which is impervious to any change in the system. In wcTD the bath variables are not dynamical variables\footnote{Dynamical variables are those which are determined consistently by the interplay between the system and the bath through their coupled equations of motion}, they only provide TD parameters such as a temperature in canonical ensemble, or, in addition, a chemical potential, in grand canonical ensemble. 

The classical thermodynamic relations among the internal energy $\mathcal{U}$, enthalpy $\mathcal{H}$, Helmholtz free energy $\mathcal{F}$ and Gibbs free energy $\mathcal{G}$ in conjunction with the temperature $T$, entropy $\mathcal{S}$, pressure $P$ and volume $\mathcal{V}$ are well-known. From the first law, 
\begin{equation}\label{U1}
	d\mathcal{U}=T\,d\mathcal{S}-P\,d\mathcal{V}\,.
\end{equation}
With $\mathcal{U}=\mathcal{U}(\mathcal{S},\mathcal{V})$, we have
\begin{align}
	T&=\biggl(\frac{\partial\mathcal{U}}{\partial\mathcal{S}}\biggr)_{\mathcal{V}}\,,&P&=-\biggl(\frac{\partial\mathcal{U}}{\partial V}\biggr)_{\mathcal{S}}\,.
\end{align}
By virtue of \eqref{U1} the enthalpy $\mathcal{H}=\mathcal{U}+P\mathcal{V}$ obeys
\begin{equation}
	d\mathcal{H}=T\,d\mathcal{S}+\mathcal{V}\,dP\,.
\end{equation}
With $\mathcal{H}=\mathcal{H}(\mathcal{S},P)$ we have
\begin{align}
	T&=\biggl(\frac{\partial\mathcal{H}}{\partial\mathcal{S}}\biggr)_{P}\,,&\mathcal{V}&=\biggl(\frac{\partial\mathcal{H}}{\partial P}\biggr)_{\mathcal{S}}\,.
\end{align}
Likewise, for the Helmholtz free energy $\mathcal{F}=\mathcal{U}-T\mathcal{S}$, we have
\begin{align}\label{E:gkhfguer}
	d\mathcal{F}&=-\mathcal{S}\,dT-P\,d\mathcal{V}\,, &&\text{whence} \,, &\mathcal{S}&=-\biggl(\frac{\partial\mathcal{F}}{\partial T}\biggr)_{\mathcal{V}}\,,&P&=-\biggl(\frac{\partial\mathcal{F}}{\partial\mathcal{V}}\biggr)_{T}\,.
\end{align}
Thus $\mathcal{F}=\mathcal{F}(T,V)$. Finally, the Gibbs free energy $\mathcal{G}=\mathcal{H}-T\mathcal{S}$ obeys
\begin{align}\label{E:rtijrsabb}
	d\mathcal{G}&=-\mathcal{S}\,dT+\mathcal{V}\,dP\,, &&\text{whence} \,,&\mathcal{S}&=-\biggl(\frac{\partial\mathcal{G}}{\partial T}\biggr)_{P}\,,&V&=\biggl(\frac{\partial\mathcal{G}}{\partial P}\biggr)_{T}\,.
\end{align}
Thus $\mathcal{G}=\mathcal{G}(T,P)$. Many more relations can be derived from these three basic relations. These relations are mutually compatible based on differential calculus.

Now we turn to the wcTD of quantum systems. (To distinguish them from strong-coupling (sc) thermodynamics  all quantities defined in the context of traditional (weak-coupling) thermodynamics are identified with a subscript $\Theta$.) The state of a quantum system in contact with a heat bath at temperature\footnote{Hereafter, we will choose the units such that $k_{B}=1$.} $T=(k_B \beta)^{-1}$ with vanishing coupling is described by the density matrix $\hat{\rho}_{s}$
\begin{equation}\label{E:rkgdfs}
	\hat{\rho}_{s}=\frac{e^{-\beta \hat{H}_{s}}}{\mathcal{Z}_{\Theta}}
\end{equation}
where 
\begin{equation} \label{ZTheta}
	\mathcal{Z}_{\Theta}=\operatorname{Tr}_{s}e^{-\beta\hat{H}_{s}}
\end{equation}
is the canonical partition function.  Here $\hat{H}_{s}$ is the Hamiltonian of the system and is assumed to be independent of the inverse temperature $\beta=T^{-1}$.  The notation $\operatorname{Tr}$ with a subscript $s$ or $b$ represents the sum over the states of the system or the bath respectively. The density matrix $\hat{\rho}_{s}$ is a time-independent Hermitian operator and is normalized to unity
\begin{equation}
	\operatorname{Tr}_{s}\hat{\rho}_{s}=1\,,
\end{equation}
to ensure unitarity.

The free energy $\mathcal{F}_{\Theta}$ of a quantum system in a canonical distribution is 
\begin{equation}\label{E:trkjndks}
	\mathcal{F}_{\Theta}=-\frac{1}{\beta}\,\ln \mathcal{Z}_{\Theta}\,.
\end{equation}
The quantum expectation value $\langle\hat{H}_{s}\rangle$ is identified with the internal energy $\mathcal{U}_{\Theta}$ of the quantum system, and can be found by
\begin{equation}\label{E:rtbkfsre}
	\mathcal{U}_{\Theta}=\langle\hat{H}_{s}\rangle=\frac{1}{\mathcal{Z}_{\Theta}}\,\operatorname{Tr}_{s}\bigl\{\hat{H}_{s}\,e^{-\beta\hat{H}_{s}}\bigr\}=-\frac{\partial}{\partial\beta}\ln\mathcal{Z}_{\Theta}=\mathcal{F}_{\Theta}+\partial_{\beta}\mathcal{F}_{\Theta}\,.
\end{equation}
Motivated by \eqref{E:gkhfguer}, we can define the entropy $\mathcal{S}_{\Theta}$ of the system by
\begin{equation}\label{E:rtbkbs}
	\mathcal{S}_{\Theta}=\beta^{2}\partial_{\beta}\mathcal{F}_{\Theta}\,,
\end{equation}
and it is connected with the free energy by the relation
\begin{equation}
	\mathcal{F}_{\Theta}=\mathcal{U}_{\Theta}-T\,\mathcal{S}_{\Theta}\,.
\end{equation}
Substituting \eqref{E:trkjndks} into \eqref{E:rtbkbs},  the entropy of the quantum system can be expressed in terms of the density matrix 
\begin{equation}
	\mathcal{S}_{\Theta}=-\operatorname{Tr}_{s}\Bigl\{\hat{\rho}_{s}\ln \hat{\rho}_{s}\Bigr\}\,,
\end{equation}
which is seen to be the von Neumann entropy. The von Neumann entropy plays an important role in quantum information as a measure of quantum entanglement, and can be used to measure the non-classical correlation in a pure-state system. (Beware of issues at zero temperature -- see the Discussions section.)

The heat capacity $\mathcal{C}_{\Theta}=\partial\mathcal{U}_{\Theta}/\partial T=-\beta^{2}\partial_{\beta}\mathcal{U}_{\Theta}$ can be expressed as
\begin{equation}\label{E:ehdfsd}
	\mathcal{C}_{\Theta}=-2\beta^{2}\partial_{\beta}\mathcal{F}_{\Theta}-\beta^{3}\partial_{\beta}\mathcal{F}_{\Theta}=-\beta\,\partial_{\beta}\mathcal{S}_{\Theta}=\beta^{2}\Bigl[\langle\hat{H}_{s}^{2}\rangle-\langle\hat{H}_{s}\rangle^{2}\Bigr]\geq0\,.
\end{equation}
Up to this point, under the vanishing system-bath coupling assumption, all the quantum thermodynamic potentials and relations still resemble their classical counterparts.

\subsection{Quantum system in a heat bath with non-vanishing coupling}

Next consider \textit{an interacting quantum system} \textbf{C} whose evolution is described by the Hamiltonian
\begin{equation}\label{E:nrjtbrs}
	\hat{H}_{c}=\hat{H}_{s}+\hat{H}_{i}+\hat{H}_{b}\,,
\end{equation}
where $\hat{H}_{s}, \hat{H}_{b}$ are the Hamiltonians of the system $\mathbf{S}$ and the bath $\mathbf{B}$, respectively and $\hat{H}_{i}$ accounts for the interaction between them. Suppose initially the composite \textbf{C}=$\mathbf{S}+\mathbf{B}$ is in a global thermal equilibrium state which is stationary, and thus has reversible dynamics,
\begin{align}
	\hat{\rho}_{c}&=\frac{e^{-\beta\hat{H}_{c}}}{\mathcal{Z}_{c}}\,,&&\text{with}&\mathcal{Z}_{c}&=\operatorname{Tr}_{sb}e^{-\beta\hat{H}_{c}}\,,
\end{align}
at the inverse temperature $\beta^{-1}$. The quantity $\mathcal{Z}_{c}$ is the partition function for the global thermal state.

In the case of vanishing coupling between the system and the bath, we may approximate the total Hamiltonian $\hat{H}_{c}$ to leading order by $\hat{H}_{c}\simeq\hat{H}_{s}+\hat{H}_{b}$. Since $[\hat{H}_{s},\hat{H}_{b}]=0$, we notice that
\begin{equation}\label{E:erbdfere}
	\frac{1}{\mathcal{Z}_{b}}\,\operatorname{Tr}_{b}e^{-\beta\hat{H}_{c}}\simeq\frac{1}{\mathcal{Z}_{b}}\,\operatorname{Tr}_{b}\bigl\{e^{-\beta\hat{H}_{s}}e^{-\beta\hat{H}_{b}}\bigr\}=e^{-\beta\hat{H}_{s}}\,,
\end{equation}
with the partition function of the free bath being given by
\begin{equation}
	\mathcal{Z}_{b}=\operatorname{Tr}_{b}e^{-\beta\hat{H}_{b}}\,.
\end{equation}
Eq.~\eqref{E:erbdfere} implies that the reduced state $\hat{\rho}_{r}=\operatorname{Tr}_{b}\hat{\rho}_{c}$, which is also stationary, will assume a canonical form
\begin{align}\label{E:ehdfd}
	\rho_{r}&=\frac{e^{-\beta \hat{H}_{s}}}{\mathcal{Z}_{s}}=\frac{1}{\mathcal{Z}_{c}}\operatorname{Tr}_{b}e^{-\beta\hat{H}_{c}}\,,&&\text{with}&\mathcal{Z}_{s}&=\operatorname{Tr}_{s}e^{-\beta\hat{H}_{s}}\,,
\end{align}
that is, $\mathcal{Z}_{c}\simeq \mathcal{Z}_{s}\mathcal{Z}_{b}$ in the limit of vanishing system-bath coupling. In addition, \eqref{E:ehdfd} ensures the proper normalization condition 
\begin{equation}
	\operatorname{Tr}_{s}\rho_{r}=1\,.
\end{equation}
Thus we have made connection with \eqref{E:rkgdfs}. That is, in the weak limit of the system-bath interaction, the reduced density matrix of the interacting composite system in the global thermal state will take the same canonical form as \eqref{E:rkgdfs}, hence to some degree  justifies the choice of the system state being \eqref{E:rkgdfs} in that particular context. Hereafter we will denote the reduced density matrix of the system by $\hat{\rho}_{s}$.

When the interaction between the system and the bath cannot be neglected, the righthand side of \eqref{E:erbdfere} no longer holds. In addition, non-commutating nature among the operators $\hat{H}_{s}$, $\hat{H}_{i}$ and $\hat{H}_{b}$ prevents us from writing 
\begin{equation}\label{E:kgbkjfsd}
	e^{-\beta(\hat{H}_{s}+\hat{H}_{i}+\hat{H}_{b})}\neq e^{-\beta(\hat{H}_{s}+\hat{H}_{b})}e^{-\beta\hat{H}_{i}}\,,
\end{equation}
due to $[\hat{H}_{s},\hat{H}_{i}]\neq0$ and $[\hat{H}_{b},\hat{H}_{i}]\neq0$ in general. In fact, according to the Baker-Campbell-Haussdorff (BCH) formula, a decomposition like \eqref{E:kgbkjfsd} will have the form
\begin{align}\label{E:erhgfdbjgs}
		e^{-\beta(\hat{H}_{s}+\hat{H}_{b})}e^{-\beta\hat{H}_{i}}&=\exp\biggl\{-\beta(\hat{H}_{s}+\hat{H}_{i}+\hat{H}_{b})+\frac{\beta^{2}}{2!}\,\bigl[\hat{H}_{s}+\hat{H}_{b},\hat{H}_{i}\bigr]\biggr.\\
		&\qquad\qquad-\biggl.\frac{\beta^{3}}{3!}\,\Bigl(\frac{1}{2}\,\bigl[\bigl[\hat{H}_{s}+\hat{H}_{b},\hat{H}_{i}\bigr],\hat{H}_{i}\bigr]+\frac{1}{2}\,\bigl[\hat{H}_{s}+\hat{H}_{b},\bigl[\hat{H}_{s}+\hat{H}_{b},\hat{H}_{i}\bigr]\bigr]\Bigr)+\cdots\biggr\}\,.\notag
\end{align}
The exponent on the righthand side typically contains an infinite number of terms. This makes algebraic manipulation of the strongly interacting system rather formidable, in contrast to its classical or quantum weak-coupling counterpart.

\paragraph{Hamiltonian of mean force}
To account for non-vanishing interactions one can introduce the \textit{Hamiltonian of mean force} $H_{s}^{*}$ for the system defined by
\begin{equation}\label{E:ejddw}
	e^{-\beta \hat{H}_{s}^{*}}\equiv\frac{1}{\mathcal{Z}_{b}}\,\operatorname{Tr}_{b}e^{-\beta \hat{H}_{c}}\,.
\end{equation}
In the limit $\hat{H}_{i}$ is negligible $\hat{H}_{s}^{*}\simeq \hat{H}_{s}^{\vphantom{*}}$; otherwise, in general $\hat{H}_{s}^{*}\neq \hat{H}_{s}^{\vphantom{*}}$. The corresponding partition function $\mathcal{Z}^{*}$ is then given by
\begin{equation}\label{E:rtjdf}
	\mathcal{Z}^{*}=\operatorname{Tr}_{s}e^{-\beta\hat{H}_{s}^{*}}=\frac{1}{\mathcal{Z}_{b}}\,\operatorname{Tr}_{sb}e^{-\beta\hat{H}_{c}}=\frac{\mathcal{Z}_{c}}{\mathcal{Z}_{b}}\,.
\end{equation}
If one followed the procedure of traditional wc thermodynamics to define the free energy as $\mathcal{F}=-\beta^{-1}\ln\mathcal{Z}$, then the total free energy $\mathcal{F}_{c}$ of the composite system can be given by a simple additive expression
\begin{equation}\label{E:tebresd}
	\mathcal{F}_{c}=\mathcal{F}^{*}+\mathcal{F}_{b}\,,
\end{equation}
with $\mathcal{F}^{*}=-\beta^{-1}\ln\mathcal{Z}^{*}$ and $\mathcal{F}_{b}=-\beta^{-1}\ln\mathcal{Z}_{b}$. Likewise, one can write the reduced density matrix $\rho_{s}$ in a form similar to \eqref{E:rkgdfs}, with the replacement of $\hat{H}_{s}$ by $\hat{H}_{s}^{*}$, 
\begin{equation}\label{E:ergdfgerw}
	\hat{\rho}_{s}=\frac{1}{\mathcal{Z}_{c}}\,\operatorname{Tr}_{b}e^{-\beta\hat{H}_{c}}=\frac{e^{-\beta\hat{H}_{s}^{*}}}{\mathcal{Z}^{*}}\,,
\end{equation}
in the hope that the conventional procedures of weak-coupling thermodynamics will follow in a way outlined in \eqref{E:rkgdfs}-\eqref{E:ehdfsd}.

{It turns out that even if we have simple expressions like $\hat{\rho}_{s}$, $\mathcal{F}_{s}$ analogous to \eqref{E:rkgdfs} and \eqref{E:trkjndks}, it does not necessarily lead to an unique, unambiguous set of thermodynamic potentials and their relations.} The culprit lies in the fact that when the interaction is non-negligible one can no long assume that the total energy of the composite system is the sum of the energy of the system and that of the bath. Ambiguities arise in the division between the system and the bath, and where to place  the energy associated with interaction. Similar ambiguities exist among the definitions of all thermodynamic potentials and thus affects the relations between them. 

Two earlier approaches to introduce the thermodynamic potentials in a strongly interacting  system in a global thermal state had been proposed: One by Gelin and Thoss \cite{GT09} for quantum systems, the other by Seifert \cite{Se16} for classical systems. We shall summarize the  Gelin and Thoss approach below and present a quantum formulation of Seifert's approach following. A recent proposal by Jarzynski \cite{JA17} for classical systems contains both approaches. We shall summarize it in the next section, and use it as a guide to work out the quantum formulation for strong coupling thermodynamics.

\subsection{Strong coupling thermodynamics according to Gelin \& Thoss}\label{S:rrnd}
The first approach, introduced by Gelin \& Thoss~\cite{GT09}, is rather intuitive, because their definitions of the internal energy and the entropy are the familiar ones in traditional thermodynamics.   They define the internal energy $\mathcal{U}_{s}$ of the (reduced) system by the quantum expectation value of the system Hamiltonian alone
\begin{equation}\label{E:rtbkdgf}
	\mathcal{U}_{s}=\operatorname{Tr}_{s}\Bigl\{\hat{\rho}_{s}\,\hat{H}_{s}\Bigr\}\,,
\end{equation}
and choose the entropy to be the von Neumann (vN) entropy $\mathcal{S}_{vN}$
\begin{equation}
	\mathcal{S}_{s}=\mathcal{S}_{vN}=-\operatorname{Tr}_{s}\Bigl\{\hat{\rho}_{s}\ln \hat{\rho}_{s}\Bigr\}\,.
\end{equation}
These are borrowed from the corresponding definitions in wcTD. 

They write the same reduced density matrix \eqref{E:ehdfd} in a slightly different representation {to highlight the difference from the wcTD case,}
\begin{align}\label{E:ejerhd}
	\hat{\rho}_{s}&=\frac{1}{\mathcal{Z}_{c}}\,e^{-\beta(\hat{H}_{s}+\hat{\Delta}_{s})}=e^{-\beta(\hat{H}_{s}+\hat{\Delta}_{s}-\mathcal{F}_{c})}\,,&&\text{with}&\mathcal{Z}_{c}&=\operatorname{Tr}_{s}e^{-\beta(\hat{H}_{s}+\hat{\Delta}_{s})}=e^{-\beta \mathcal{F}_{c}}
\end{align}
where $\hat{\Delta}_{s}$ depends only on the system variables but includes all of the influence from the bath from their  interaction. Comparing this with \eqref{E:ejddw}, we note that $\hat{\Delta}_{s}$ is formally related to the Hamiltonian of mean force by
\begin{align}\label{E:tnkrjsfsa}
	e^{-\beta(\hat{H}_{s}+\hat{\Delta}_{s})}&=\frac{\mathcal{Z}_{c}}{\mathcal{Z}^{*}}\,e^{-\beta \hat{H}_{s}^{*}}=\mathcal{Z}_{b}\,e^{-\beta\hat{H}_{s}^{*}}\,,&&\Rightarrow&\hat{\Delta}_{s}&=\hat{H}_{s}^{*}-\hat{H}_{s}+\mathcal{F}_{b}\,.
\end{align} 
Finally, they let the partition function of the system take on the value $\mathcal{Z}_{c}$, which is distinct from $\mathcal{Z}^{*}$. Thus the corresponding free energy will be given by $\mathcal{F}_{c}$ which contains all the contributions from the composite \textbf{C}.

Although in this approach the definitions of internal energy and entropy of the system are quite intuitive, these two thermodynamic quantities do not enjoy simple relations with the partition function $\mathcal{Z}_{c}$, as in \eqref{E:rtbkfsre} and \eqref{E:rtbkbs}.  From \eqref{E:ejerhd}, we can show\footnote{Here some discretion is advised in taking the derivative with respect to $\beta$ because in general an operator will not commute with its own derivative. See details in App.~\ref{S:opera}.}
\begin{align}
	-\frac{\partial}{\partial\beta}\,\ln\mathcal{Z}_{c}=-\frac{1}{\mathcal{Z}_{c}}\frac{\partial}{\partial\beta}\operatorname{Tr}_{s}e^{-\beta(\hat{H}_{s}+\hat{\Delta}_{s})}&=\langle \hat{H}_{s}\rangle+\langle\hat{\Delta}_{s}\rangle+\beta\,\langle\partial_{\beta}\hat{\Delta}_{s}\rangle\,,\label{E:lpernere}
\end{align}
that is, 
\begin{equation}
	\mathcal{U}_{s}\neq-\frac{\partial}{\partial\beta}\,\ln\mathcal{Z}_{c}\,,
\end{equation}
and
\begin{equation}\label{E:dfbesffs}
	\mathcal{F}_{c}=\mathcal{U}_{s}+\langle\hat{\Delta}_{s}\rangle+\beta\,\langle\partial_{\beta}\hat{\Delta}_{s}\rangle-\beta\,\partial_{\beta}\mathcal{F}_{c}\,.
\end{equation}
Here $\langle\cdots\rangle$ represents the expectation value taken with respect to the density matrix $\hat{\rho}_{c}$ of the composite. For a system operator $\hat{O}_{s}$, this definition yields an expectation value equal to that with respect to the reduced density matrix $\hat{\rho}_{s}$, namely, 
\begin{equation}
	\langle\hat{O}_{s}\rangle_{s}=\operatorname{Tr}_{s}\Bigl\{\hat{\rho}_{s}\,\hat{O}_{s}\Bigr\}=\operatorname{Tr}_{sb}\Bigl\{\hat{\rho}_{c}\,\hat{O}_{s}\Bigr\}=\langle\hat{O}_{s}\rangle\,.
\end{equation}
Likewise, the von Neumann entropy $\mathcal{S}_{vN}$ can be expressed in terms of the free energy $\mathcal{F}_{c}$ by
\begin{equation}
	\mathcal{S}_{vN}=\beta\operatorname{Tr}_{s}\Bigl\{\hat{\rho}_{s}\bigl(\hat{H}_{s}+\hat{\Delta}_{s}-\mathcal{F}_{c}\bigr)\Bigr\}=\beta\langle \hat{H}_{s}\rangle+\beta\langle\hat{\Delta}_{s}\rangle-\beta\,\mathcal{F}_{c}=\beta^{2}\partial_{\beta}\mathcal{F}_{c}-\beta^{2}\langle\partial_{\beta}\hat{\Delta}_{s}\rangle\,,
\end{equation}
which does not look like \eqref{E:rtbkbs}. Additionally, we observe the entropy so defined is not additive, that is,
\begin{equation}
	\mathcal{S}_{vN}+\mathcal{S}_{b}=-\operatorname{Tr}_{s}\Bigl\{\hat{\rho}_{s}\ln \hat{\rho}_{s}\Bigr\}-\operatorname{Tr}_{b}\Bigl\{\hat{\rho}_{s}\ln \hat{\rho}_{b}\Bigr\}\neq-\operatorname{Tr}_{sb}\Bigl\{\hat{\rho}_{c}\ln \hat{\rho}_{c}\Bigr\}=\mathcal{S}_{c}\,.
\end{equation}
Here $\mathcal{S}_{c}$ and $\mathcal{S}_{b}$ are the von Neumann entropies of the composite and the free bath, respectively. {Note that} the $\hat{\rho}_{b}$ in this formulation is {the density matrix of the free bath,} not the reduced density matrix of the bath, namely,
\begin{equation} 
	\hat{\rho}_{b}\neq\operatorname{Tr}_{s}\hat{\rho}_{c}\,.
\end{equation}
The reduced density matrix of the bath will contain an additional overlap with the system from their coupling.

When the internal energy of the system given by the expectation value of the system Hamiltonian \eqref{E:rtbkdgf}, the specific heat $\mathcal{C}_{s}$ will take the form, with the help of \eqref{E:dfbesffs},
\begin{align}\label{E:rthrjts}
	\mathcal{C}_{s}=-\beta^{2}\partial_{\beta}\langle\hat{H}_{s}\rangle&=-\beta\,\partial_{\beta}\mathcal{S}_{vN}-\beta^{2}\Bigl[\langle\partial_{\beta}\hat{\Delta}_{s}\rangle-\partial_{\beta}\langle\hat{\Delta}_{s}\rangle\Bigr]\,.
\end{align}
In general $\langle\partial_{\beta}\hat{\Delta}_{s}\rangle\neq\partial_{\beta}\langle\hat{\Delta}_{s}\rangle$ since the reduced density matrix $\hat{\rho}_{s}$ also has a {temperature} dependence. We thus see in this case the heat capacity cannot be directly given as the derivative of the (von Neumann) entropy with respect to $\beta$, as in \eqref{E:ehdfsd}.

In short, in the G\&T formulation the thermodynamic potentials of the system are defined in a direct and intuitive way, without explicit reference to the bath or the composite, except that the partition function $\mathcal{Z}_{c}$ is still needed to bridge the relevant relations among these potentials. G\&T introduce an operator $\hat{\Delta}_{s}$ to highlight the foreseen ambiguity when the system is strongly coupled with the bath. From \eqref{E:rtjdf} and \eqref{E:tnkrjsfsa}, we see, formally 
\begin{align}\label{E:ottote}
	e^{-\beta(\hat{H}_{s}+\hat{\Delta}_{s})}&=\operatorname{Tr}_{b}e^{-\beta(\hat{H}_{s}+\hat{H}_{i}+\hat{H}_{b})}\,,&&\Rightarrow&\hat{\Delta}_{s}&=-\beta^{-1}\ln\operatorname{Tr}_{b}e^{-\beta(\hat{H}_{s}+\hat{H}_{i}+\hat{H}_{b})}-\hat{H}_{s}\,.
\end{align}
In the limit of weak coupling, $\hat{H}_{i}\approx0$, \eqref{E:ottote} reduces to
\begin{align}\label{E:ottotwe}
	\hat{\Delta}_{s}&\approx-\beta^{-1}\ln\operatorname{Tr}_{b}e^{-\beta(\hat{H}_{s}+\hat{H}_{b})}-\hat{H}_{s}=-\beta^{-1}\ln \mathcal{Z}_{b}\,.
\end{align}
Hence in this limit, $\hat{\Delta}_{s}$ reduces to a $c$-number and plays the role of the free energy $\mathcal{F}_{b}$ of the free bath. This can also be seen from \eqref{E:tnkrjsfsa} since $\hat{H}_{s}^{*}\approx\hat{H}_{s}$ in the same limit. Observe $\hat{\Delta}_{s}\approx\mathcal{F}_{b}$ in the weak coupling limit annuls the expression in the square brackets in \eqref{E:rthrjts} and  restores the traditional relation \eqref{E:ehdfsd} between the heat capacity and the entropy. However, even in the weak coupling limit, the internal energy still cannot be given by \eqref{E:rtbkfsre}. The disparity lies in the identification of $\mathcal{Z}_{c}$ as the partition function of the system. As is clearly seen from \eqref{E:lpernere}, in the weak coupling limit, we have
\begin{align}
	-\frac{\partial}{\partial\beta}\,\ln\mathcal{Z}_{c}\approx-\frac{1}{\mathcal{Z}_{c}}\frac{\partial}{\partial\beta}\operatorname{Tr}_{s}\Bigl\{e^{-\beta\hat{H}_{s}}\mathcal{Z}_{b}\Bigr\}=\langle\hat{H}_{s}\rangle_{s}+\langle\hat{H}_{b}\rangle_{b}\,.
\end{align}
This implies that $\mathcal{Z}_{c}$ is not a good candidate for the partition function of the system. A more suitable option would be $\mathcal{Z}_{c}/\mathcal{Z}_{b}$.

\subsection{Quantum formulation of Seifert's thermodynamics at strong coupling}\label{S:jrtbsgsgd}

If we literally follow \eqref{E:ejddw} and identify $\hat{H}_{s}^{*}$ as the effective Hamiltonian operator of the (reduced) system, we will nominally interpret that the reduced system assumes a canonical distribution. Thus it is natural to identify $\mathcal{Z}^{*}$ as the partition function associated with the reduced state of the system.

Suppose we maintain the thermodynamic relations regardless of the coupling strength between the system and the bath. From \eqref{E:rtbkfsre} and \eqref{E:rtbkbs}, we will arrive at expressions of the internal energy and entropy of the system. This is essentially Seifert's approach ~\cite{Se16} to the thermodynamics at strong coupling for classical systems. Here we will present the quantum-mechanical version of it. First, from \eqref{E:ejddw}, we have the explicit form of the Hamiltonian of mean force $\hat{H}_{s}^{*}$
\begin{align}\label{E:ebrehsd}
	\hat{H}_{s}^{*}=-\beta^{-1}\ln\operatorname{Tr}_{b}\Bigl\{\exp\Bigl [-\beta \hat{H}_{s}-\beta \hat{H}_{i}-\beta\bigl(\hat{H}_{b}-\mathcal{F}_{b}\bigr)\Bigr]\Bigr\}\,.
\end{align}
This is the operator form of $\mathcal{H}(\xi_{s},\lambda)$ in Eq.~(5) of \cite{Se16}. Noting the non-commutative characters of the operators. Since $[\hat{H}_{s},\hat{H}_{i}]\neq0$, 
\begin{equation}
	e^{-\beta(\hat{H}_{s}+\hat{H}_{i}+\hat{H}_{b})}\not\to e^{-\beta\hat{H}_{s}}e^{-\beta(\hat{H}_{i}+\hat{H}_{b})}\,.
\end{equation}
If one prefers to factor out  $e^{-\beta\hat{H}_{s}}$ from $e^{-\beta(\hat{H}_{s}+\hat{H}_{i}+\hat{H}_{b})}$, one can use the BCH formula, outlined in App.~\ref{S:opera}, to expand out the operator products to a certain order commensurate with a specified degree of accuracy.

Second, it is readily seen that $p^{\text{eq}}(\xi_{s}\vert\lambda)$ in Eq.~(4) of \cite{Se16} is the reduced density matrix $\hat{\rho}_{s}$ of the system \eqref{E:ergdfgerw}. The (Helmholtz) free energy $\mathcal{F}$ in Seifert's Eq~(7)  is exactly the free energy of the reduced system $\mathcal{F}^{*}$ in \eqref{E:rtjdf}.

With these identifications, it is easier to {find the rest of the physical} quantities in Seifert's strong coupling thermodynamics. We now proceed to derive  the entropy and the internal energy, i.e., Eqs.~(8), (9) of \cite{Se16}, for quantum systems in his framework.  From the thermodynamic relation \eqref{E:rtbkbs}, we have
\begin{align}
	\mathcal{S}_{s}=\beta^{2}\frac{\partial\mathcal{F}^{*}}{\partial\beta}&=-\beta\mathcal{F}^{*}+\beta\,\operatorname{Tr}_{s}\biggl\{\hat{\rho}_{s}\Bigl(\hat{H}_{s}^{*}+\beta\,\partial_{\beta}\hat{H}_{s}^{*}\Bigr)\biggr\}\,.\label{E:dfkekrfd}
\end{align}
Here we recall that even though the operators $\hat{H}_{s}^{*}$ and $\partial_{\beta}\hat{H}_{s}^{*}$ in general do not commute, the trace operation allowing for cyclic permutations of the operator products eases the difficulties in their manipulation. 
Since \eqref{E:ergdfgerw} implies the operator identity
\begin{equation}
	\beta \hat{H}_{s}^{*}=\beta\mathcal{F}^{*}-\ln\hat{\rho}_{s}\,,
\end{equation}
we can recast \eqref{E:dfkekrfd} to
\begin{align}
	\mathcal{S}_{s}=\operatorname{Tr}_{s}\bigl\{\hat{\rho}_{s}\bigl(-\ln\rho_{s}+\beta^{2}\partial_{\beta}\hat{H}_{s}^{*}\bigr)\bigr\}=\mathcal{S}_{vN}+\beta^{2}\langle\partial_{\beta}H_{s}^{*}\rangle\neq-\operatorname{Tr}_{s}\Bigl\{\hat{\rho}_{s}\ln\hat{\rho}_{s}\Bigr\}=\mathcal{S}_{vN}\,.\label{E:derbdf}
\end{align} 
This is the quantum counterpart of Seifert's entropy, Eq.~(8) of \cite{Se16}.  This entropy is often called the `thermodynamic' entropy in the literature. Note that it is not equal to the von Neumann (`statistical') entropy $\mathcal{S}_{vN}$ of the system.

The internal energy can be given by the thermodynamic relation 
\begin{equation}\label{E:ebdjerdf}
	\mathcal{U}_{s}=\mathcal{F}^{*}+\beta^{-1}\mathcal{S}_{s}\,.
\end{equation}
Thus from \eqref{E:dfkekrfd}, we obtain, 
\begin{align}\label{E:rthrndf}
	\mathcal{U}_{s}=\operatorname{Tr}_{s}\bigl\{\hat{\rho}_{s}\bigl(\hat{H}_{s}^{*}+\beta\,\partial_{\beta}\hat{H}_{s}^{*}\bigr)\bigr\}=\langle\hat{H}_{s}^{*}\rangle+\beta\,\langle\partial_{\beta}\hat{H}_{s}^{*}\rangle\neq\langle\hat{H}_{s}^{*}\rangle\,.
\end{align}
This deviation results from the fact that $\hat{H}_{s}^{*}$, introduced in \eqref{E:ejddw} may depend on $\beta$. When we take this into consideration, we can also verify that the internal energy can also be consistently given by Eq.~\eqref{E:rtbkfsre}
\begin{align}\label{E:dfbjhs}
	-\frac{\partial}{\partial\beta}\,\ln\mathcal{Z}^{*}&=\frac{1}{\mathcal{Z}^{*}}\operatorname{Tr}_{s}\Bigl\{\Bigl(\hat{H}_{s}^{*}+\beta\,\partial_{\beta}\hat{H}_{s}^{*}\Bigr)\,e^{-\beta\hat{H}_{s}^{*}}\Bigr\}=\mathcal{U}_{s}\,.
\end{align}
In fact, we can also show, by recognizing $\mathcal{Z}^{*}=\mathcal{Z}_{c}/\mathcal{Z}_{b}$, that
\begin{equation}\label{E:edfhjer}
	\mathcal{U}_{s}=\langle\hat{H}_{s}\rangle+\Bigl [\langle\hat{H}_{i}\rangle+\langle\hat{H}_{b}\rangle-\langle\hat{H}_{b}\rangle_{b}\Bigr]\neq\langle\hat{H}_{s}\rangle\,, 
\end{equation}
with
\begin{align}
	\langle\hat{H}_{b}\rangle_{b}&\equiv\operatorname{Tr}_{b}\bigl\{\hat{\rho}_{b}\,\hat{H}_{b}\bigr\}\,,&\langle\hat{H}_{b}\rangle&\equiv\operatorname{Tr}_{sb}\bigl\{\hat{\rho}_{c}\,\hat{H}_{b}\bigr\}\,,&\langle\hat{H}_{s}\rangle&\equiv\operatorname{Tr}_{sb}\bigl\{\hat{\rho}_{c}\,\hat{H}_{s}\bigr\}=\langle\hat{H}_{s}\rangle_{s}\,.
\end{align}
Eq.~\eqref{E:edfhjer} implies that the internal energy, defined by \eqref{E:dfbjhs}, accommodates more than mere $\langle\hat{H}_{s}\rangle_{s}$. The additional pieces contain contributions from the bath and the interaction. In particular, when the coupling between the system and the bath is not negligible, we have $\langle\hat{H}_{b}\rangle\neq\langle\hat{H}_{b}\rangle_{b}$ in general. In fact, even the internal energy defined in \eqref{E:rtbkdgf} in the first (G\&T) approach also has influence from the bath because the reduced density matrix $\hat{\rho}_{s}$ includes all the effects of the bath on the system.

So far, we have encountered three possible definitions of internal energies, namely,  $\langle\hat{H}_{s}\rangle$, $\langle\hat{H}_{s}^{*}\rangle$, and $\mathcal{U}_{s}$. As can be seen from \eqref{E:rthrndf} and \eqref{E:edfhjer}, essentially they differ by the amount of the bath and the interaction energy which are counted toward the system energy. This ambiguity clearly arises from strong coupling between the system and the bath. When the system-bath interaction is negligibly small, we have $\langle\hat{H}_{i}\rangle\approx0$, and\footnote{In the weak coupling limit, the full density matrix of the composite  is approximately given by the product of that of the system and the bath.} $\langle\hat{H}_{b}\rangle\approx\langle\hat{H}_{b}\rangle_{b}$, and these three energies become equivalent.

To explicate the physical meaning of $\langle\hat{H}_{s}^{*}\rangle$, we note that from \eqref{E:ergdfgerw}, we can write $\langle\hat{H}_{s}^{*}\rangle$ as
\begin{align}
	\langle\hat{H}_{s}^{*}\rangle&=-\frac{1}{\beta}\,\operatorname{Tr}_{s}\bigl\{\hat{\rho}_{s}\ln\hat{\rho}_{s}\bigr\}+\mathcal{F}^{*}=\beta^{-1}\mathcal{S}_{vN}+\mathcal{F}^{*}\,,&&\text{or}&\langle\hat{H}_{s}^{*}\rangle&=\mathcal{F}^{*}+\beta^{-1}\,\mathcal{S}_{vN}\,.
\end{align}
This offers an interesting comparison with \eqref{E:ebdjerdf}, where $\mathcal{F}^{*}=\mathcal{U}_{s}-\beta^{-1}\mathcal{S}_{s}$. It may appear that we can replace the pair $(\mathcal{U}_{s},\mathcal{S}_{s})$ by another pair $(\langle\hat{H}_{s}^{*}\rangle,\mathcal{S}_{vN})$, leaving $\mathcal{F}^{*}$ unchanged, thus suggesting an alternative definition of internal energy by $\langle\hat{H}_{s}^{*}\rangle$ and that of entropy by $\mathcal{S}_{vN}$. However, in so doing, the new energy and entropy will not satisfy a simple thermodynamic relation like \eqref{E:rtbkfsre} and \eqref{E:rtbkbs}. This is a good sign,  as it is an indication that certain internal consistency exists in the choice of the thermodynamic variables.

We now investigate the differences between the two definitions of entropy. From \eqref{E:derbdf}, we obtain
\begin{align}
	T\bigl(\mathcal{S}_{s}-\mathcal{S}_{vN}\bigr)=\operatorname{Tr}_{s}\Bigl\{\hat{\rho}_{s}\bigl(\beta\,\partial_{\beta}\hat{H}_{s}^{*}\bigr)\Bigr\}=\beta\,\partial_{\beta}\operatorname{Tr}_{s}\Bigl\{\hat{\rho}_{s}\,\hat{H}_{s}^{*}\Bigr\}-\beta\operatorname{Tr}_{s}\Bigl\{\bigl(\partial_{\beta}\hat{\rho}_{s}\bigr)\hat{H}_{s}^{*}\Bigr\}\,.
\end{align}
The factor $\partial_{\beta}\hat{\rho}_{s}$ can be written as
\begin{align}
	\partial_{\beta}\hat{\rho}_{s}=\partial_{\beta}\biggl[\frac{1}{\mathcal{Z}_{c}}\operatorname{Tr}_{b}e^{-\beta\hat{H}_{c}}\biggr]&=\langle\hat{H}_{c}\rangle\,\hat{\rho}_{s}-\operatorname{Tr}_{b}\bigl\{\hat{\rho}_{c}\,\hat{H}_{c}\bigr\}
\end{align}
with $\partial_{\beta}\mathcal{Z}_{c}=-\langle\hat{H}_{c}\rangle\,\mathcal{Z}_{c}$. We then obtain
\begin{align}\label{E:trirhnh}
	T\bigl(\mathcal{S}_{s}-\mathcal{S}_{vN}\bigr)=\beta\,\partial_{\beta}\langle\hat{H}_{s}^{*}\rangle+\beta\Bigl [\langle\hat{H}_{c}\,\hat{H}_{s}^{*}\rangle-\langle\hat{H}_{c}\rangle\langle\hat{H}_{s}^{*}\rangle\Bigr]\,.
\end{align}
Thus,  part of the difference between the two entropies result from the correlation between the full Hamiltonian $\hat{H}_{c}$ and the Hamiltonian of mean force $\hat{H}_{s}^{*}$. This correlation will disappear in the vanishing coupling limit because there is no interaction to bridge the system and the bath. We also note that in the same limit, $\langle\hat{H}_{s}^{*}\rangle\approx\langle\hat{H}_{s}\rangle$ becomes $\beta$-independent, and both definitions of the entropy turn synonymous.

Since the von Neumann entropy $\mathcal{S}_{vN}$ can be used as a measure of entanglement between the system and the bath, we often introduce the quantum mutual information $I_{sb}$ to quantify how they are correlated,
\begin{equation}
	I_{sb}=\mathcal{S}_{vN}+\mathcal{S}'_{b}-\mathcal{S}_{c}\geq0\,,
\end{equation}
where $\mathcal{S}'_{b}$ is the von Neumann entropy associated with the reduced density matrix $\hat{\varrho}_{b}$ of the bath, in contrast to $\mathcal{S}_{b}$ we have met earlier. This mutual information can be related to the quantum relative entropy $S(\hat{\rho}_{c}\Vert\hat{\rho}_{s}\otimes\hat{\varrho}_{b})$ by
\begin{equation}
	S(\hat{\rho}_{c}\Vert\hat{\rho}_{s}\otimes\hat{\varrho}_{b})=\operatorname{Tr}_{sb}\Bigl\{\hat{\rho}_{c}\ln\hat{\rho}_{c}-\hat{\rho}_{c}\ln\hat{\rho}_{s}\otimes\hat{\varrho}_{b}\Bigr\}=I_{sb}\,,
\end{equation}
because $\hat{\varrho}_{b}=\operatorname{Tr}_{s}\hat{\rho}_{c}$. On the other hand, Eqs.~\eqref{E:tebresd} and \eqref{E:dfkekrfd} imply that the {thermodynamic} entropy $\mathcal{S}_{s}$ is additive
\begin{equation}\label{E:dbrtdaa}
	\mathcal{S}_{s}+\mathcal{S}_{b}=\mathcal{S}_{c}\,,
\end{equation}
from which we find
\begin{equation}
	I_{sb}=\bigl(\mathcal{S}'_{b}-\mathcal{S}_{b}\bigr)+\bigl(\mathcal{S}_{vN}-\mathcal{S}_{s}\bigr)\,.
\end{equation}
This and \eqref{E:trirhnh} provide different perspectives on how the difference between the two system entropies is related to the system-bath entanglement, and how the system-bath coupling has a role in establishing such correlations.

Following the definitions of the internal energy \eqref{E:dfbjhs} and the entropy \eqref{E:dfkekrfd}, the heat capacity of the system still satisfies a familiar relation
\begin{equation}
	\mathcal{C}_{s}=-\beta^{2}\,\partial_{\beta}\mathcal{U}_{s}=\beta^{2}\,\partial_{\beta}^{2}\ln\mathcal{Z}^{*}=-\beta\,\partial_{\beta}\mathcal{S}_{s}\,.
\end{equation}
Compared with \eqref{E:rthrjts}, with the help of \eqref{E:edfhjer},  we clearly see their difference given by
\begin{align}
	-\beta^{2}\,\partial_{\beta}\bigl(\mathcal{U}_{s}-\langle\hat{H}_{s}\rangle\bigr)&=-\beta^{2}\,\partial_{\beta}\Bigl [\langle\hat{H}_{i}\rangle+\langle\hat{H}_{b}\rangle-\langle\hat{H}_{b}\rangle_{b}\Bigr].
\end{align}

\subsection{Issues}
Both quantum formulations for thermodynamics at strong coupling are based on plausible assumptions and are mathematically sound. In the first (G\&T) approach, one starts with intuitive definitions of the thermodynamics quantities, inspired by traditional thermodynamics for classical systems premised on vanishingly weak coupling between the system and the bath. This leads to modifications in the thermodynamic relations of the relevant thermodynamics quantities. In the second (Seifert) approach, one opts to maintain the familiar thermodynamic relations but is compelled to deal with rather obscure interpretation of the thermodynamic potentials.

To understand their strength and weakness more explicitly we can apply these two methods to a simple and completely solvable model, namely, a Brownian oscillator linearly but strongly coupled with a large  (or infinitely large, as modeled by a scalar field) bath. We will see both approaches at some point or other produce ambiguous or paradoxical results. We make a few observations in the following.

\subsubsection{Entropy}

1) It has been discussed in~\cite{NA02, OC06, HB08} that the von Neumann entropy $\mathcal{S}_{vN}$ will not approach to zero for the finite system-bath coupling in the limit of zero temperature, but the thermodynamic entropy $\mathcal{S}_{s}$, defined in Approach I\!I, behaves nicely in the same limit. 

2) It has been {shown}~\cite{HZ95} that if the composite is in a global thermal state the discrete energy spectrum of the undamped oscillator will become a continuous one with a unique ground level. This supports physics described by the thermodynamic entropy $\mathcal{S}_{s}$. 

3) It has been argued~\cite{JB04, HL09,HB08} that the entanglement between the system and the bath prevents the von Neumann entropy from approaching zero. Since the reduced state of the system remains mixed, the projective measurement of the system Hamiltonian operator $\hat{H}_{s}$ indicates that{when the density matrix is expanded in terms of the unperturbed states of the system,} the reduced system can still be in the exited states even at zero temperature. Without quantum entanglement between the system and the bath, the lowest energy level of the composite system will be given by the tensor product of the ground state of the unperturbed system and bath, that is, a pure state. In this case, the von Neumann entropy will goes to zero as expected, and this is the scenario occurred in traditional thermodynamics in the vanishing system-bath coupling limit.

\subsubsection{Internal Energy}

4) It has been discussed~\cite{HA11,HI06,HI08,IH09} that the internal energy defined in Approach I\!I can lead to anomalous behavior of the heat capacity in the low temperature limit. When the system, consisting of a quantum oscillator~\cite{HA11} or a free particle~\cite{HI06,HI08,IH09} is coupled to a heat bath modeled by a large number of quantum harmonic oscillators, the heat capacity of the system can become negative if the temperature of the bath is sufficiently low. If the internal energy defined in Approach I is used to compute the heat capacity, then it has been shown that the heat capacity remains positive for all nonzero temperature but vanishes in the zero bath temperature limit, for a system with one harmonic oscillator~\cite{HI06}, or a finite number of coupled harmonic oscillators~\cite{HC17}. {This discrepancy may result from the fact that the internal energy defined in Approach I\!I contains contributions from the interaction and the bath Hamiltonian.}

\section{Jarzynski's Thermodynamics at strong coupling for classical systems}\label{S:tbwjrjc}

Since a quantum formulation of Jarzynski's thermodynamics at strong coupling is our goal, and to make the presentation self-contained, it is useful to briefly summarize the main points in his approach for classical systems.
Consider a composite \textbf{C} comprising of a system $\mathbf{S}$ of interest interacting with  a heat bath $\mathbf{B}$,  with phase space variables $x$, $y$ respectively.   The total Hamiltonian $H_{c}$ of the composite is given by
\begin{equation}
	H_{c}(x,y)=H_{s}(x)+H_{i}(x,y)+H_{b}(y)\,,
\end{equation}
where $H_{i}$ is the interaction Hamiltonian which can be of arbitrary strength. Suppose the bath of finite volume $V_{b}$, which can depend on the state of the bath, is subjected to a constant external pressure $P$. Assume that initially this composite system exerted by the same constant pressure $P$ is in thermal equilibrium at temperature $\beta^{-1}$, described by the distribution
\begin{align}\label{E:rutbgf}
	\rho(x,y)&=\frac{e^{-\beta[H_{c}(x,y)+PV_{b}(y)]}}{\mathcal{Z}_{c}}\,,&\mathcal{Z}_{c}&=\int\!dx\,dy\;e^{-\beta[H_{c}(x,y)+PV_{b}(y)]}\,.
\end{align}
We also define the counterparts of \eqref{E:rutbgf} for the bath $\mathbf{B}$,
\begin{align}
	\rho_{b}(y)&=\frac{e^{-\beta[H_{b}(y)+PV_{b}(y)]}}{\mathcal{Z}_{b}}\,,&\mathcal{Z}_{b}&=\int\!dy\;e^{-\beta[H_{b}(y)+PV_{b}(y)]}\,.
\end{align}
Note that both $H_{c}$ and $H_{b}$ are independent of $\beta$ and $P$. Introduce a new thermodynamical potential $\phi(x;P,\beta)$ by
\begin{equation}\label{E:gnftewefs}
	\phi(x;P,\beta)=-\beta^{-1}\,\ln\frac{\displaystyle\int\!dy\;e^{-\beta(H_{i}+H_{b}+PV_{b})}}{\displaystyle\int\!dy\;e^{-\beta(H_{b}+PV_{b})}}=G_{i}(x;P,\beta)-G_{b}(P,\beta)\,,
\end{equation}
where the Gibbs free energy $G$ is defined by
\begin{align}\label{E:rnrtr}
	e^{-\beta\,G_{i}(x;P,\beta)}&=\mathcal{Z}_{i}(x;P,\beta)=\int\!dy\;e^{-\beta(H_{i}+H_{b}+PV_{b})}\,,\\
	e^{-\beta\,G_{b}(P,\beta)}&=\mathcal{Z}_{b}(P,\beta)=\int\!dy\;e^{-\beta(H_{b}+PV_{b})}\,.
\end{align}
Note that $G_{b}=\mathcal{G}_{b}$ which was defined in \eqref{E:rtijrsabb} since it does not depend on the micro states of the bath.

Classically, $G_{i}(x;P,\beta)$ can be viewed as the micro-state Gibbs free energy of the bath $\mathbf{B}$ when it is driven by the system $\mathbf{S}$ in a fixed state $x$. Thus $\phi$ is the free energy difference of the bath  due to the intervention of the system $\mathbf{S}$ in its state $x$. Equivalently we can write \eqref{E:gnftewefs} as
\begin{align}
	e^{-\beta\phi}&=\frac{\mathcal{Z}_{i}}{\mathcal{Z}_{b}}\,,
\end{align}
where $\mathcal{Z}_{i}(x;P,\beta)$ is the partition function of the bath when it is driven by the system in a fixed state $x$.

The reduced state of the system is given by
\begin{equation}
	\rho_{s}(x;P,\beta)=\int\!dy\;\rho_{c}(x,y)=\frac{1}{\mathcal{Z}}_{c}\int\!dy\,e^{-\beta(H_{c}+PV_{b})}=\frac{e^{-\beta(H_{s}+\phi)}}{\mathcal{Z}_{s}}=\frac{\mathcal{Z}_{i}}{\mathcal{Z}_{c}}\,e^{-\beta\,H_{s}}\,,
\end{equation}
where we have defined the partition function $\mathcal{Z}_{s}$ of the system by
\begin{align}\label{E:rtkjrgs}
	\mathcal{Z}_{s}&=\frac{\mathcal{Z}_{c}}{\mathcal{Z}_{b}}=e^{-\beta \mathcal{G}_{s}}=\int\!dx\;e^{-\beta(H_{s}+\phi)}\,,&\mathcal{G}_{s}&=\mathcal{G}_{c}-\mathcal{G}_{b}=-\beta^{-1}\ln\int\!dx\;e^{-\beta(H_{s}+\phi)}\,,
\end{align}
and the Gibbs free energy $\mathcal{G}_{c}$ of the composite is
\begin{equation}
	\mathcal{G}_{c}=-\beta^{-1}\ln\mathcal{Z}_{c}\,.
\end{equation}
The concept of dynamical volume $V$ can be introduced via $\phi(x;P,\beta)$. There are two distinct forms depending on the way we define this volume.

\subsection{`Bare' representation}

In this representation, one focuses on the system \textbf{S}  in a fashion similar to the traditional weak-coupling thermodynamics. Manipulations of thermodynamic formulas are performed on the system variables only. Thus, it is formulated more or less in parallel with the first approach introduced in Sec.~\ref{S:rrnd}. Later we will address their dissimilarities.

The dynamical volume of the system can be defined by
\begin{equation}\label{E:ngrjkgsd}
	V_{s}^{(b)}(x;P,\beta)=\frac{\phi(x;P,\beta)}{P}\,.
\end{equation}
The superscript $(b)$ of variables is used to remind us that they are in the bare representation. We define the micro-state internal energy $U_{s}(x)$ and enthalpy $\mathfrak{H}_{s}(x;P,\beta)$ by
\begin{align}\label{E:truhbgd}
	U^{(b)}_{s}(x)&=H_{s}(x)\,,&\mathfrak{H}^{(b)}_{s}(x;P,\beta)&=U^{(b)}_{s}(x)+P\,V^{(b)}_{s}(x;P,\beta)=U^{(b)}_{s}(x)+\phi(x;P,\beta)\,.
\end{align}
Thus the reduced state can also be expressed with
\begin{align}\label{E:htbsdx}
	\rho_{s}(x;P,\beta)&=\frac{e^{-\beta\,\mathfrak{H}^{(b)}_{s}(x;P,\beta)}}{\mathcal{Z}_{s}}=e^{-\beta\,(\mathfrak{H}^{(b)}_{s}-\mathcal{G}^{(b)}_{s})}\,,&&\text{with}&\mathcal{Z}_{s}(P,\beta)&=\int\!dx\;e^{-\beta\,\mathfrak{H}^{(b)}_{s}(x;P,\beta)}\,,
\end{align}
or
\begin{align}
	\mathcal{G}_{s}(P,\beta)&=-\beta^{-1}\ln\int\!dx\;e^{-\beta\,\mathfrak{H}^{(b)}_{s}}\,.
\end{align}
The ensemble averages of $V_{s}^{(b)}$, $U_{s}^{(b)}$, $\mathfrak{H}_{s}^{(b)}$ are then given by
\begin{align}
	\mathcal{V}_{s}^{(b)}(P,\beta)&=\int\!dx\;\rho_{s}(x;P,\beta)\,V_{s}^{(b)}(x;P,\beta)\,,\label{E:irtjri1}\\
	\mathcal{U}_{s}^{(b)}(P,\beta)&=\int\!dx\;\rho_{s}(x;P,\beta)\,U_{s}^{(b)}(x)\,,\\
	\mathcal{H}_{s}^{(b)}(P,\beta)&=\int\!dx\;\rho_{s}(x;P,\beta)\,\mathfrak{H}_{s}^{(b)}(x;P,\beta)\,.\label{E:irtjri3}
\end{align}
Eqs.~\eqref{E:truhbgd} and \eqref{E:irtjri1}--\eqref{E:irtjri3} imply
\begin{equation}
	\mathcal{H}_{s}^{(b)}(P,\beta)=\mathcal{U}_{s}^{(b)}(P,\beta)+P\,\mathcal{V}_{s}^{(b)}(P,\beta)\,.
\end{equation}
We define the entropy $\mathcal{S}^{(b)}_{s}$ by
\begin{equation}\label{E:lfpmz} 
	\mathcal{S}^{(b)}_{s}(P,\beta)=-\int\!dx\;\rho_{s}(x;P,\beta)\,\ln\rho_{s}(x;P,\beta)\,.
\end{equation}
Substituting \eqref{E:htbsdx} into \eqref{E:lfpmz} leads to
\begin{align}
	\mathcal{S}^{(b)}_{s}(P,\beta)=-\int\!dx\;\rho_{s}\Bigl[-\beta\,\mathfrak{H}^{(b)}_{s}(x;P,\beta)+\beta\,\mathcal{G}_{s}(P,\beta)\Bigr]=\beta\Bigl[\mathcal{H}_{s}^{(b)}(P,\beta)-\mathcal{G}_{s}(P,\beta)\Bigr]\,,
\end{align}
that is
\begin{equation}
	\mathcal{G}_{s}(P,\beta)=\mathcal{H}_{s}^{(b)}(P,\beta)-\beta^{-1}\,\mathcal{S}^{(b)}_{s}(P,\beta)\,.
\end{equation}
Note with these definitions, we have
\begin{align}
	\mathcal{S}^{(b)}_{s}&\neq\beta^{2}\,\frac{\partial}{\partial\beta}\Bigl[-\beta^{-1}\ln \mathcal{Z}_{s}\Bigr]=\beta^{2}\,\frac{\partial}{\partial\beta}\,\mathcal{G}_{s}^{(b)}\,,&\mathcal{U}_{s}^{(b)}&\neq-\frac{\partial}{\partial\beta}\ln \mathcal{Z}_{s}=\frac{\partial}{\partial\beta}\Bigl[\beta\,\mathcal{G}^{(b)}_{s}\Bigr]\,.
\end{align}
The main feature of the bare representation is that we only deal with the variables of the system exclusively, without any reference to those of the composite or the bath.

\subsection{`Partial molar' representation}\label{S:tbrgf}
In this representation, the description of the system is essentially given in terms of the composite {and the bath}, although in the intermediate manipulations we still need some information of the reduced state of the system.

An alternative way to define the dynamical volume of the system $\mathbf{S}$ is
\begin{equation}\label{E:gnfjgn}
	V^{(p)}_{s}(x;P,\beta)=\frac{\partial}{\partial P}\,\phi(x;P,\beta)\,,
\end{equation}
which is in general different from $V^{(b)}_{s}$, defined in \eqref{E:ngrjkgsd}. Eq.~\eqref{E:gnfjgn} can be written as
\begin{align}
	V^{(p)}_{s}(x;P,\beta)=-\beta^{-1}\frac{\partial}{\partial P}\,\ln\frac{\mathcal{Z}_{i}}{\mathcal{Z}_{b}}=V^{(p)}_{i}(x;P,\beta)-V^{(p)}_{b}(P,\beta)\,,\label{E:gnkjes}
\end{align}
where 
\begin{align}
	V^{(p)}_{i}(x;P,\beta)&=\frac{1}{\mathcal{Z}_{i}}\int\!dy\;V_{b}(y;P,\beta)\,e^{-\beta(H_{i}+H_{b}+PV_{b})}\,,\label{E:rrbgssdf1}\\
	V^{(p)}_{b}(P,\beta)&=\frac{1}{\mathcal{Z}_{b}}\int\!dy\;V_{b}(y;P,\beta)\,e^{-\beta(H_{b}+PV_{b})}=\mathcal{V}^{(p)}_{b}(P,\beta)\,.\label{E:rrbgssdf2}
\end{align}
Thus $V^{(p)}_{s}$ can be roughly understood as the change of the volume of the bath due to the influence of the system in its micro state $x$. The corresponding average is then given by
\begin{align}
	\mathcal{V}^{(p)}_{s}(P,\beta)&=\int\!dx\;\rho_{s}(x;P,\beta)\,V^{(p)}_{i}(x;P,\beta)-\int\!dx\;\rho_{s}(x;P,\beta)\,\mathcal{V}^{(p)}_{b}(P,\beta)\notag\\
	&=\frac{1}{\mathcal{Z}_{c}}\int\!dx\,dy\;V_{b}\,e^{-\beta(H_{s}+H_{i}+H_{b}+PV_{b})}-\mathcal{V}^{(p)}_{b}(P,\beta)\label{E:prtojgnf}\\
	&=\mathcal{V}^{(p)}_{c}(P,\beta)-\mathcal{V}^{(p)}_{b}(P,\beta)\,,\label{E:grbgkbs}
\end{align}
where we have invoked an useful identity
\begin{equation}
	\frac{\mathcal{Z}_{i}}{\mathcal{Z}_{b}}\,e^{-\beta\,H_{s}}=e^{-\beta\,(H_{s}+\phi)}\,,
\end{equation}
and defined the average volume $\mathcal{V}^{(p)}_{c}$ of the composite by
\begin{equation}
	\mathcal{V}^{(p)}_{c}(P,\beta)=\int\!dx\;\rho_{s}(x;P,\beta)\,V^{(p)}_{i}(x;P,\beta)=\frac{1}{\mathcal{Z}_{c}}\int\!dx\,dy\;V_{b}\,e^{-\beta(H_{s}+H_{i}+H_{b}+PV_{b})}\,.
\end{equation}
The interpretation of $\mathcal{V}^{(p)}_{s}(P,\beta)$ is different from that of $V^{(p)}_{s}(x;P,\beta)$. Eq.~\eqref{E:grbgkbs} highlights the fact that the averaged dynamical volume of the system can be expressed as the difference between the averaged dynamical volume of the composite $\mathbf{C}$  and that of the bath $\mathbf{B}$ in the absence of the system. Note that in general we have $\mathcal{V}^{(p)}_{b}(P,\beta)\neq V_{b}(y;P,\beta)$. Derivations of \eqref{E:grbgkbs} also tell us that{we can identify}
\begin{align}
	\mathcal{V}^{(p)}_{c}(P,\beta)&=-\frac{1}{\beta}\frac{\partial}{\partial P}\ln \mathcal{Z}_{c}\,,&\mathcal{V}^{(p)}_{b}(P,\beta)&=-\frac{1}{\beta}\frac{\partial}{\partial P}\ln \mathcal{Z}_{b}\,,\label{E:tirgfn2}
\end{align}
which imply
\begin{align}
	\mathcal{V}^{(p)}_{c}(P,\beta)&=\frac{\partial \mathcal{G}_{c}}{\partial P}\,,&\mathcal{V}^{(p)}_{b}(P,\beta)&=\frac{\partial \mathcal{G}_{b}}{\partial P}\,,&&\Rightarrow&\mathcal{V}^{(p)}_{s}(P,\beta)=\frac{\partial \mathcal{G}_{s}}{\partial P}\,,
\end{align}
by \eqref{E:rtkjrgs}. This complies with the standard thermodynamic relation \eqref{E:rtijrsabb} which relates volume with the Gibbs free energy. 

Hinted by \eqref{E:prtojgnf}, we introduce the micro-state internal energies by 
\begin{align}
	U^{(p)}_{c}(x;P,\beta)&=\frac{1}{\mathcal{Z}_{i}}\int\!dy\;\bigl(H_{s}+H_{i}+H_{b}\bigr)\,e^{-\beta(H_{i}+H_{b}+P\,V_{b})}\,,\label{E:tbgd1}\\
	U^{(p)}_{b}(P,\beta)&=\frac{1}{\mathcal{Z}_{b}}\int\!dy\;H_{b}\,e^{-\beta(H_{b}+P\,V_{b})}=\mathcal{U}^{(p)}_{b}(P,\beta)\,.\label{E:tbgd2}
\end{align}
The micro-state internal energy $U_{s}$ of the system $\mathbf{S}$ is then defined by
\begin{equation}
	U^{(p)}_{s}(x;P,\beta)=U^{(p)}_{c}(x;P,\beta)-U^{(p)}_{b}(P,\beta)\,,
\end{equation}
and the corresponding macro averaged value is given by
\begin{align}
	\mathcal{U}^{(p)}_{s}(P,\beta)&=\int\!dx\;\rho_{s}(x;P,\beta)\,U^{(p)}_{s}(x;P,\beta)=\mathcal{U}^{(p)}_{c}(P,\beta)-\mathcal{U}^{(p)}_{b}(P,\beta)\,,\label{E:rtrfndeure}
\end{align}
where 
\begin{align}
	\mathcal{U}^{(p)}_{c}(P,\beta)&=\frac{1}{\mathcal{Z}_{c}}\int\!dx\,dy\;\bigl(H_{s}+H_{i}+H_{b}\bigr)\,e^{-\beta(H_{s}+H_{i}+H_{b}+P\,V_{b})}\,,
\end{align}
and $\mathcal{U}^{(p)}_{b}(P,\beta)$ are the mean energies of the composite system and the free bath.

Likewise, from \eqref{E:tbgd1}, \eqref{E:tbgd2} and \eqref{E:rrbgssdf1}, \eqref{E:rrbgssdf2}, {the definitions of the internal energy and volume lead us to} define the micro-state enthalpy $\mathfrak{H}^{(p)}_{c}(x;P,\beta)$ of the composite $\mathbf{C}$, and the counterpart $\mathfrak{H}^{(p)}_{b}(P,\beta)$ of the bath $\mathbf{B}$ by
\begin{align}
	\mathfrak{H}^{(p)}_{c}(x;P,\beta)&=U^{(p)}_{c}(x;P,\beta)+P\,V^{(p)}_{i}(x;P,\beta)\notag\\
	&=\frac{1}{\mathcal{Z}_{i}}\int\!dy\;\bigl(H_{s}+H_{i}+H_{b}+P\,V_{b}\bigr)\,e^{-\beta(H_{i}+H_{b}+P\,V_{b})}\,,\\
	\mathfrak{H}^{(p)}_{b}(P,\beta)&=U^{(p)}_{b}(P,\beta)+P\,V^{(p)}_{b}(P,\beta)\notag\\
	&=\frac{1}{\mathcal{Z}_{b}}\int\!dy\;\bigl(H_{b}+P\,V_{b}\bigr)\,e^{-\beta(H_{b}+P\,V_{b})}=\mathcal{H}^{(p)}_{b}(P,\beta)\,.
\end{align}
The micro-state enthalpy $\mathfrak{H}^{(p)}_{s}(x;P,\beta)$ of the system $\mathbf{S}$ is given by
\begin{equation}
	\mathfrak{H}^{(p)}_{s}(x;P,\beta)=\mathfrak{H}^{(p)}_{c}(x;P,\beta)-\mathfrak{H}^{(p)}_{b}(P,\beta)\,,
\end{equation}
and then its average is
\begin{align}
	\mathcal{H}^{(p)}_{s}(P,\beta)=\int\!dx\;\rho_{s}(x;P,\beta)\,\mathfrak{H}^{(p)}_{s}(x;P,\beta)&=-\frac{\partial}{\partial\beta}\,\ln\mathcal{Z}_{s}=\mathcal{H}^{(p)}_{c}(P,\beta)-\mathcal{H}^{(p)}_{b}(P,\beta)\,,
\end{align}
where
\begin{align}
	\mathcal{H}^{(p)}_{c}(P,\beta)&=-\frac{\partial}{\partial\beta}\,\ln\mathcal{Z}_{c}=\frac{1}{\mathcal{Z}_{c}}\int\!dx\,dy\;\bigl(H_{s}+H_{i}+H_{b}+P\,V_{b}\bigr)\,e^{-\beta(H_{s}+H_{i}+H_{b}+P\,V_{b})}\,,\\
	\mathcal{H}^{(p)}_{b}(P,\beta)&=-\frac{\partial}{\partial\beta}\,\ln\mathcal{Z}_{b}=\frac{1}{\mathcal{Z}_{b}}\int\!dy\;\bigl(H_{b}+P\,V_{b}\bigr)\,e^{-\beta(H_{b}+P\,V_{b})}\,.
\end{align}
Thus the enthalpy of the system $\mathbf{S}$ is again expressed by the difference between the enthalpy of the composite and the bath. We can  conclude {consistently} with the following relations
\begin{align}
	\mathcal{H}^{(p)}_{c}(P,\beta)&=\mathcal{U}^{(p)}_{c}(P,\beta)+P\,\mathcal{V}^{(p)}_{c}(P,\beta)\,,&\mathcal{H}^{(p)}_{s}(P,\beta)&=\mathcal{U}^{(p)}_{s}(P,\beta)+P\,\mathcal{V}^{(p)}_{s}(P,\beta)\,,\\
	\mathcal{H}^{(p)}_{b}(P,\beta)&=\mathcal{U}^{(p)}_{b}(P,\beta)+P\,\mathcal{V}^{(p)}_{b}(P,\beta)\,.
\end{align}
Moreover, since the Gibbs free energy and the enthalpy of the composite system are
\begin{align}
	\mathcal{G}_{c}&=-\beta^{-1}\ln\mathcal{Z}_{c}\,,&\mathcal{H}^{(p)}_{c}&=-\beta^{-1}\frac{\partial}{\partial\beta}\,\ln\mathcal{Z}_{c}\,,
\end{align}
we find
\begin{align}
	\beta\bigl(\mathcal{H}^{(p)}_{c}-\mathcal{G}_{c}\bigr)&=\beta^{2}\,\frac{\partial \mathcal{G}_{c}}{\partial\beta}=-\int\!dx\,dy\;\rho_{c}\,\ln\rho_{c}\,.
\end{align}
Thus{it motivates us to} define the entropy $\mathcal{S}^{(p)}_{c}$ of the composite  by
\begin{equation}
	\mathcal{S}^{(p)}_{c}=-\int\!dx\,dy\;\rho_{c}\,\ln\rho_{c}=\beta\bigl(\mathcal{H}^{(p)}_{c}-\mathcal{G}_{c}\bigr)=\beta^{2}\,\frac{\partial \mathcal{G}_{c}}{\partial\beta}\,,
\end{equation}
and the same arguments lead to the entropy $\mathcal{S}^{(p)}_{b}$ of the bath by
\begin{equation}
	\mathcal{S}^{(p)}_{b}=-\int\!dy\;\rho_{b}\,\ln\rho_{b}=\beta\bigl(\mathcal{H}^{(p)}_{b}-\mathcal{G}_{b}\bigr)=\beta^{2}\,\frac{\partial \mathcal{G}_{b}}{\partial\beta}\,.
\end{equation}
This enables us to define the the entropy $\mathcal{S}^{(p)}_{s}$ of the system $\mathbf{S}$ by
\begin{equation}
	\mathcal{S}^{(p)}_{s}=\mathcal{S}^{(p)}_{c}-\mathcal{S}^{(p)}_{b}=\beta\bigl(\mathcal{H}^{(p)}_{s}-\mathcal{G}_{s}\bigr)=\beta^{2}\,\frac{\partial \mathcal{G}_{s}}{\partial\beta}.
\end{equation}
Note, however, 
\begin{equation}
	\mathcal{S}^{(p)}_{s}=-\int\!dx\,dy\;\rho\,\ln\rho+\int\!dy\;\rho_{b}\,\ln\rho_{b}\neq-\int\!dx\;\rho_{s}\,\ln\rho_{s}=\mathcal{S}^{(p)}_{vN}\,,
\end{equation}
which is different from the Gibbs entropy of the system, or the system entropy in the bare representation.

In contrast to the bare representation, in the partial molar representation the variables of the system are almost always defined in terms of the difference between their counterparts of the composite and the bath. The definitions of the thermodynamic potentials are introduced in a similar spirit as those in the second approach outlined in Sec.~\ref{S:jrtbsgsgd}. {Thus these definitions preserve the thermodynamic relations used in the wcTD with the caveat that the interpretation or definition of the corresponding micro-state quantities can be obscure at times.}

\section{Quantum formulation of Jarzynski's strong coupling thermodynamics}\label{S:tubfgdgfd}

We now provide a quantum formulation of Jarzynski's classical results \cite{JA17}.  The Hamiltonian operator of the composite $\mathbf{C}=\mathbf{S} +\mathbf{B}$ is assumed to take the form
\begin{equation}
	\hat{H}_{c}=\hat{H}_{s}+\hat{H}_{i}+\hat{H}_{b}+J\cdot\hat{A}_{b}\,,
\end{equation}
where $J$ is some external $c$-number drive acting on the bath via a bath operator $\hat{A}_{b}$.

If the composite system is in thermal equilibrium, its state is described by the generalized canonical ensemble and the corresponding density matrix is
\begin{align}
	\hat{\rho}_{c}&=\frac{1}{\mathcal{Z}_{c}}\,e^{-\beta\,\hat{H}_{c}}\,,&\mathcal{Z}_{c}&=\operatorname{Tr}_{sb}\Bigl\{e^{-\beta\,\hat{H}_{c}}\Bigr\}\,,
\end{align}
where $\mathcal{Z}_{c}$, a $c$-number, is the partition function of the composite. For later convenience, we also define the corresponding quantities for the bath $\mathbf{B}$ when it is coupled to the system $\mathbf{S}$,
\begin{align}
	\hat{\rho}_{b}&=\frac{1}{\mathcal{Z}_{b}}\,e^{-\beta\,(\hat{H}_{b}+J\cdot\hat{A}_{b})}\,,&\mathcal{Z}_{b}&=\operatorname{Tr}_{b}\Bigl\{e^{-\beta\,(\hat{H}_{b}+J\cdot\hat{A}_{b})}\Bigr\}\,.
\end{align}
We introduce the Hamiltonian operator of mean force $\hat{H}_{s}^{*}$, as before, by
\begin{equation}
	e^{-\beta\,\hat{H}_{s}^{*}}\equiv\frac{1}{\mathcal{Z}_{b}}\,\operatorname{Tr}_{b}\Bigl\{e^{-\beta\,\hat{H}_{c}}\Bigr\}\,,
\end{equation}
such that the reduced density matrix of the system $\mathbf{S}$ takes the form
\begin{align}
	\hat{\rho}_{s}&\equiv\operatorname{Tr}_{b}\hat{\rho}_{c}=\frac{1}{\mathcal{Z}_{s}}\,e^{-\beta\,\hat{H}_{s}^{*}}\,,&&\text{with}&\mathcal{Z}_{s}&=\frac{\mathcal{Z}_{c}}{\mathcal{Z}_{b}}=\operatorname{Tr}_{s}\Bigl\{e^{-\beta\,\hat{H}_{s}^{*}}\Bigr\}\,.
\end{align}
The quantity $\mathcal{Z}_{s}$ can be viewed as an effective partition function of the system $\mathbf{S}$. This is motivated by the observation that, in the absence of coupling between \textbf{S} and \textbf{B},  or in the weak coupling limit, the composite is additive so its partition function is the product of those of the subsystems, i.e., $\mathcal{Z}_{c}=\mathcal{Z}_{s}\mathcal{Z}_{b}$. The difference $\hat{H}_{s}^{*}-\hat{H}_{s}$  modifies the dynamics of the system $\mathbf{S}$ due to its interaction with the bath $\mathbf{B}$. 

In fact, by the construction, $e^{-\beta\,H_{s}^{*}}$, once sandwiched by the appropriate states of the system $\mathbf{S}$ and expressed in the imaginary-time path integral formalism, is formally $e^{-S_{\textsc{cg}}}$, where $S_{\textsc{cg}}$ is the coarse-grained effective action of the system $\mathbf{S}$, wick-rotated to the imaginary time. Thus formally $\beta(\hat{H}_{s}^{*}-\hat{H}_{s})$ is equivalent to the influence action in the imaginary time formalism.

Similar to the classical formulations, we may have two different representations of the operator $\hat{A}_{s}$ of the system.

\subsection{`Bare' representation}
In the bare representation, we may define
\begin{equation}
	\hat{A}_{s}=\frac{\hat{H}_{s}^{*}-\hat{H}_{s}}{J}\,,
\end{equation}
and the internal energy operator $\hat{U}_{s}$ and the  enthalpy operator $\hat{\mathfrak{H}}_{s}$, respectively,  by
\begin{align}
	\hat{U}_{s}&=\hat{H}_{s}\,,&\hat{\mathfrak{H}}_{s}&=\hat{H}_{s}^{*}\,,
\end{align}
with expectation values given by
\begin{align}
	\mathcal{U}_{s}&=\operatorname{Tr}_{s}\Bigl\{\hat{\rho}_{s}\,\hat{U}_{s}\Bigr\}\,,&\mathcal{H}_{s}&=\operatorname{Tr}_{s}\Bigl\{\hat{\rho}_{s}\,\hat{\mathfrak{H}}_{s}\Bigr\}=\mathcal{U}_{s}+J\cdot\mathcal{A}_{s}\,,
\end{align}
corresponding to the internal energy and the enthalpy we are familiar with, respectively, where
\begin{equation}
	\mathcal{A}_{s}=\operatorname{Tr}_{s}\Bigl\{\hat{\rho}_{s}\,\hat{A}_{s}\Bigr\}\,.
\end{equation}
The entropy is chosen to be the von Neumann entropy of the system
\begin{equation}
	\mathcal{S}_{s}=\operatorname{Tr}_{s}\Bigl\{\hat{\rho}_{s}\,\ln\hat{\rho}_{s}\Bigr\}=\beta\bigl(\mathcal{H}_{s}-\mathcal{G}_{s}\bigr)\,.
\end{equation}
These definitions are in exact parallel to those in the classical formulation contained in \eqref{E:ngrjkgsd}--\eqref{E:lfpmz}.

\subsection{`Partial molar' representation}

In this representation, in an analogy with Sec.~\ref{S:tbrgf}, for the system $\mathbf{S}$, we can alternatively define the operator $\hat{A}_{s}(x)$ that corresponds to $\hat{A}_{b}(y)$ of the bath $\mathbf{B}$ by
\begin{equation}
	\hat{A}_{s}(x)=\frac{\partial}{\partial J}\bigl(\hat{H}_{s}^{*}-\hat{H}_{s}\bigr)=\frac{\partial\hat{H}_{s}^{*}}{\partial J}\,.
\end{equation}
The last equality results from the fact that $\hat{H}_{s}$ has no dependence on the external parameter $J$. Owing  to the non-commutativity of operators the micro-physics interpretation of the operator $\hat{A}_{s}(x)$ is not so transparent. We first focus on its quantum expectation value $\mathcal{A}_{s}$
\begin{align}
	\mathcal{A}_{s}=\operatorname{Tr}_{s}\Bigl\{\hat{\rho}_{s}\,\hat{A}_{s}\Bigr\}=\operatorname{Tr}_{s}\Bigl\{\hat{\rho}_{s}\,\frac{\partial\hat{H}_{s}^{*}}{\partial J}\Bigr\}=\frac{1}{\mathcal{Z}_{s}}\operatorname{Tr}_{s}\Bigl\{e^{-\beta\,\hat{H}^{*}_{s}}\,\frac{\partial\hat{H}_{s}^{*}}{\partial J}\Bigr\}\,.\label{E:etijfgdfg}
\end{align}
From \eqref{E:tbgfbsd}, the righthand side of \eqref{E:etijfgdfg} can be identified as
\begin{equation}
	\operatorname{Tr}_{s}\Bigl\{e^{-\beta\,\hat{H}^{*}_{s}}\,\frac{\partial\hat{H}_{s}^{*}}{\partial J}\Bigr\}=-\beta^{-1}\,\frac{\partial}{\partial J}\operatorname{Tr}_{s}\Bigl\{e^{-\beta\,\hat{H}^{*}_{s}}\Bigr\}\,,
\end{equation}
and thus we have
\begin{equation}
	\mathcal{A}_{s}=-\beta^{-1}\frac{\partial}{\partial J}\ln\mathcal{Z}_{s}\,.
\end{equation}
The advantage of this expression is that the observation of $\mathcal{Z}_{s}=\mathcal{Z}_{c}/\mathcal{Z}_{b}$ enables us to write $\mathcal{A}_{s}$ as
\begin{equation}\label{E:potrjsfs}
	\mathcal{A}_{s}=-\beta^{-1}\frac{\partial}{\partial J}\ln\frac{\mathcal{Z}_{c}}{\mathcal{Z}_{b}}=-\beta^{-1}\frac{\partial}{\partial J}\ln\mathcal{Z}_{c}+\beta^{-1}\frac{\partial}{\partial J}\ln\mathcal{Z}_{b}=\mathcal{A}_{c}-\mathcal{A}_{b}\,,
\end{equation}
where we have defined the corresponding expectation values for the composite $\mathbf{C}$ and the bath $\mathbf{B}$ by
\begin{align}
	\mathcal{A}_{c}&=-\beta^{-1}\frac{\partial}{\partial J}\ln\mathcal{Z}_{c}\,,&\mathcal{A}_{b}&=-\beta^{-1}\frac{\partial}{\partial J}\ln\mathcal{Z}_{b}\,.
\end{align}
In particular we can check that
\begin{align}
	\mathcal{A}_{b}&=-\beta^{-1}\frac{\partial}{\partial J}\ln\operatorname{Tr}_{b}\Bigl\{e^{-\beta\,(\hat{H}_{b}+J\cdot\hat{A}_{b})}\Bigr\}=\frac{1}{\mathcal{Z}_{b}}\operatorname{Tr}_{b}\Bigl\{e^{-\beta\,(\hat{H}_{b}+J\cdot\hat{A}_{b})}\;\hat{A}_{b}\,\Bigr\}=\operatorname{Tr}_{b}\Bigl\{\hat{\rho}_{b}\,\hat{A}_{b}\,\Bigr\}\,,\label{E:kfgkfke}
\end{align}
that is, $\mathcal{A}_{b}$ indeed is the expectation value of the operator $\hat{A}_{b}$. We also note that \eqref{E:kfgkfke} can be written as
\begin{equation}
	\mathcal{A}_{b}=\operatorname{Tr}_{sb}\Bigl\{\hat{\rho}_{b}\,\hat{A}_{b}\,\Bigr\}\,.
\end{equation}
This can nicely bridge with $\mathcal{A}_{c}$ for the composite,
\begin{equation}\label{E:ntrkjbs}
	\mathcal{A}_{c}=\operatorname{Tr}_{sb}\Bigl\{\hat{\rho}\,\hat{A}_{b}\,\Bigr\}\,.
\end{equation}
As seen in \eqref{E:potrjsfs} the expectation value $\mathcal{A}$ is additive, that is, its value for the combined systems is equal to the sum of those of the subsystems, $\mathcal{A}_{c}=\mathcal{A}_{s}+\mathcal{A}_{b}$. In fact, this additive property holds for all the thermodynamics potentials introduced afterwards. This is an important feature in Jarzynski's partial molar representation or in Seifert's approach.

From this aspect, we can interpret $\mathcal{A}_{s}$ as the change of $\mathcal{A}_{b}$ due to the intervention of the system $\mathbf{S}$. For example, 
consider a photon gas inside a cavity box, one side of which is a movable classical mirror {and is exerted by a constant pressure}. Assume originally the photon gas and the mirror are in thermal equilibrium. In this cavity we now  place a Brownian charged oscillator and maintain the new composite system in  thermal equilibrium at the same temperature{and the same pressure}\footnote{The equilibration process in this example can be awfully complicated if we mind the subtleties regarding whether the photon gas can ever reach thermal equilibrium in a cavity whose walls are perfectly reflective and so on. For the present argument we assume equilibration is possible and there is no leakage of the photons.}. Then we should note that there is a minute change in the mean position of the mirror before and after the Brownian charged oscillator is placed into the cavity. This change can also be translated to an effective or dynamical size of the charged oscillator due to its interaction with the photon gas.

From this example, it is tempting to identify $J\cdot\hat{A}_{s}$ as some quantum work operator\footnote{Its value depends on the interaction between the system and the bath and when this interaction is switch on. It is thus path-dependent in the parameter space of the coupling constant.}.  {Alternatively we may} view it or its expectation as some additional ``energy content'' of the system $\mathbf{S}$ due to its interaction with the bath when the composite is acted upon  by an external agent $J$, since $\hat{A}_{s}$ is related to $\hat{H}_{s}^{*}-\hat{H}_{s}$~\cite{CH05}. Inspired by this observation and taking the hint from \eqref{E:potrjsfs}, we introduce the enthalpy of the system $\mathbf{S}$ by
\begin{equation}
	\mathcal{H}_{s}=-\frac{\partial}{\partial\beta}\ln\mathcal{Z}_{s}=-\frac{\partial}{\partial\beta}\ln\mathcal{Z}_{c}+\frac{\partial}{\partial\beta}\ln\mathcal{Z}_{b}=\mathcal{H}_{c}-\mathcal{H}_{b}\,,
\end{equation}
where we have identified the enthalpies of the composite $\mathbf{C}=\mathbf{S}+\mathbf{B}$ and the bath $\mathbf{B}$ as
\begin{align}
	\mathcal{H}_{c}&=-\frac{\partial}{\partial\beta}\ln\mathcal{Z}_{c}\,,&\mathcal{H}_{b}&=-\frac{\partial}{\partial\beta}\ln\mathcal{Z}_{b}\,.
\end{align}
We rewrite them and obtain
\begin{align}
	\mathcal{H}_{c}&=\langle\hat{H}_{s}\rangle+\langle\hat{H}_{i}\rangle+\langle\hat{H}_{b}\rangle+J\cdot\mathcal{A}\,,\label{E:tnhgg1}\\
	\mathcal{H}_{b}&=\langle\hat{H}_{b}\rangle_{b}+J\cdot\mathcal{A}_{b}\,.\label{E:tnhgg2}
\end{align}
This implies that 1)
\begin{align}
	\mathcal{H}_{s}=\mathcal{H}_{c}-\mathcal{H}_{b}&=\Bigl[\langle\hat{H}_{s}\rangle+\langle\hat{H}_{i}\rangle+\langle\hat{H}_{b}\rangle-\langle\hat{H}_{b}\rangle_{b}\Bigr]+J\cdot\mathcal{A}_{s}\,,
\end{align}
and 2) the internal energy $\mathcal{U}_{s}$ of the system $\mathbf{S}$ can be consistently defined by
\begin{equation}\label{E:rthrjf}
	\mathcal{U}_{s}=\langle\hat{H}_{s}\rangle+\langle\hat{H}_{i}\rangle+\Bigl[\langle\hat{H}_{b}\rangle-\langle\hat{H}_{b}\rangle_{b}\Bigr]\,.
\end{equation}
In fact this is the same internal energy \eqref{E:edfhjer} obtained in Seifert's approach. From \eqref{E:tnhgg1} and \eqref{E:tnhgg2}, we can also define the internal energy of the composite system and of the bath by
\begin{align}
	\mathcal{U}_{c}&=\langle\hat{H}_{s}\rangle+\langle\hat{H}_{i}\rangle+\langle\hat{H}_{b}\rangle\,,\label{E:tuhsssr1}\\
	\mathcal{U}_{b}&=\langle\hat{H}_{b}\rangle_{b}\,,\label{E:tuhsssr2}
\end{align}
and thus we also conclude
\begin{equation}\label{E:rngkjfs}
	\mathcal{U}_{s}=\mathcal{U}-\mathcal{U}_{b}\,.
\end{equation}
We can see that the internal energy $\mathcal{U}_{s}$ also includes contributions that na\"ively we will not ordinarily attribute to the system, such as $\langle\hat{H}_{b}\rangle-\langle\hat{H}_{b}\rangle_{b}$. Doing so will complicate the physical connotation of the internal energy of the system.

\paragraph{Enthalpy and Energy Operators: Caution --}
In fact, we may deduce the operator form of the quantities introduced earlier. For example, we may intuitively define the enthalpy operator $\hat{\mathfrak{H}}$ of the composite 
\begin{equation}
	\hat{\mathfrak{H}}_{c}=\hat{H}_{s}+\hat{H}_{i}+\hat{H}_{b}+J\cdot\hat{A}_{b}\,,
\end{equation}
and then it is clear to see that the expectation value $\mathcal{H}_{c}$ is related to this operator by
\begin{equation}
	\mathcal{H}_{c}=\operatorname{Tr}_{sb}\Bigl\{\hat{\rho}_{c}\,\hat{\mathfrak{H}}_{c}\Bigr\}=\langle\hat{\mathfrak{H}}_{c}\rangle\,.
\end{equation}
Likewise the enthalpy operator $\hat{\mathfrak{H}}_{b}$ of the bath $\mathbf{B}$ can be define by
\begin{equation}
	\hat{\mathfrak{H}}_{b}=\hat{H}_{b}+J\cdot\hat{A}_{b}\,,
\end{equation}
and its expectation value gives $\mathcal{H}_{b}$, obtained in \eqref{E:tnhgg2},
\begin{equation}
	\mathcal{H}_{b}=\operatorname{Tr}_{b}\Bigl\{\hat{\rho}_{b}\,\hat{\mathfrak{H}}_{b}\Bigr\}=\langle\hat{\mathfrak{H}}_{b}\rangle_{b}\,.
\end{equation}
Moreover, the internal energy operator $\hat{U}_{c}$ of the composite system and the expectation value can be chosen such that
\begin{align}\label{E:rtdnsq1}
	\hat{U}_{c}&=\hat{H}_{s}+\hat{H}_{i}+\hat{H}_{b}\,,&\mathcal{U}_{c}&=\operatorname{Tr}_{sb}\Bigl\{\hat{\rho}_{c}\,\hat{U}_{c}\Bigr\}=\langle\hat{U}_{c}\rangle\,,
\end{align}
as has been given by \eqref{E:tuhsssr1}. For the bath, the internal energy operator $\hat{U}_{b}$ is, intuitively,  
\begin{equation}\label{E:rtdnsqw}
	\hat{U}_{b}=\hat{H}_{b}\,,
\end{equation}
with expectation values 
\begin{equation}
	\mathcal{U}_{b}=\operatorname{Tr}_{b}\Bigl\{\hat{\rho}_{b}\,\hat{U}_{b}\Bigr\}=\langle\hat{U}_{b}\rangle_{b}\,,
\end{equation}
consistent with \eqref{E:tuhsssr2}. Despite their intuitively appealing appearances these operator forms of the enthalpies and internal energies are not very useful. Inadvertent use of them may result in errors. For example, we cannot define the enthalpy operator of system $\mathbf{S}$ simply by the difference of $\hat{\mathfrak{H}}_{c}$ and $\hat{\mathfrak{H}}_{b}$, since
\begin{align}\label{E:rtbrsd}
	\hat{\mathfrak{H}}_{s}\stackrel{?}{=}\hat{\mathfrak{H}}_{c}-\hat{\mathfrak{H}}_{b}=\hat{H}_{s}+\hat{H}_{i}\,.
\end{align}
This result in \eqref{E:rtbrsd} is nonsensical because 1) the righthand side still explicitly depends on the bath, 2) we cannot take its trace with respect to the state of the system, $\hat{\rho}_{s}$, and thus 3) the expectation value will not be $\mathcal{H}_{s}$. This is because the operators defined this way  act on Hilbert spaces {different from that of $\hat{\rho}_{s}$}: $\hat{\mathfrak{H}}_{c}$ is an operator in the Hilbert space of the composite while $\hat{\mathfrak{H}}_{b}$ is an operator in the Hilbert space of the bath. Neither operator acts exclusively in the Hilbert space of the system. Thus extreme care is needed when manipulating the operator forms of the thermodynamical potentials. What one needs to do is to seek the local forms of these operators, i.e., operators which act only on the Hilbert space of the system. This can be done in parallel to Jarzynski's classical formulation.

\paragraph{System Enthalpy Operator: Approved --}
We first inspect the internal energy operator. Since the averaged internal energy of the composite system is given by
\begin{align}
	\mathcal{U}_{c}=\frac{1}{\mathcal{Z}_{c}}\operatorname{Tr}_{sb}\Bigl\{e^{-\beta\,\hat{H}_{c}}\,\bigl(\hat{H}_{s}+\hat{H}_{i}+\hat{H}_{b}\bigr)\Bigr\}\,,
\end{align}
we can rewrite the expressions inside the traces into
\begin{align}
	\mathcal{U}_{c}&=\operatorname{Tr}_{s}\biggl\{\frac{e^{-\beta\,\hat{H}_{s}^{*}}}{\mathcal{Z}_{s}}\,\frac{e^{+\beta\,\hat{H}_{s}^{*}}}{\mathcal{Z}_{b}}\,\operatorname{Tr}_{b}\Bigl[e^{-\beta\,\hat{H}}_{c}\,\bigl(\hat{H}_{s}+\hat{H}_{i}+\hat{H}_{b}\bigr)\Bigr]\biggr\}\notag\\
	&=\operatorname{Tr}_{s}\biggl\{\hat{\rho}_{s}\,\hat{Z}_{i}^{-1}\,e^{+\beta\,\hat{H}_{s}}\,\operatorname{Tr}_{b}\Bigl[e^{-\beta\,\hat{H}_{c}}\,\bigl(\hat{H}_{s}+\hat{H}_{i}+\hat{H}_{b}\bigr)\Bigr]\biggr\}\notag\\
	&=\operatorname{Tr}_{s}\biggl\{\hat{\rho}_{s}\,\hat{Z}_{i}^{-1}\operatorname{Tr}_{b}\Bigl[e^{+\beta\,\hat{H}_{s}}e^{-\beta\,\hat{H}_{c}}\,\bigl(\hat{H}_{s}+\hat{H}_{i}+\hat{H}_{b}\bigr)\Bigr]\biggr\}
\end{align}
where we have used the fact that $\mathcal{Z}_{c}=\mathcal{Z}_{s}\mathcal{Z}_{b}$ and the identity for the operator $\hat{Z}_{i}$
\begin{align}
	\hat{Z}_{i}&\equiv e^{+\beta\,\hat{H}_{s}}\,\operatorname{Tr}_{b}\Bigl\{e^{-\beta\,\bigl(\hat{H}_{s}+\hat{H}_{i}+\hat{H}_{b}+J\cdot\hat{A}_{b}\bigr)}\Bigr\}=\mathcal{Z}_{b}\,e^{+\beta\,\hat{H}_{s}}e^{-\beta\,\hat{H}_{s}^{*}}\,,&&\Leftrightarrow&\frac{e^{+\beta\,\hat{H}_{s}^{*}}}{\mathcal{Z}_{b}}&=\hat{Z}_{i}^{-1}\,e^{+\beta\,\hat{H}_{s}}\,.
\end{align}
If we define an internal energy operator $\hat{U}_{i}$ by
\begin{equation}
	\hat{U}_{i}=\hat{Z}_{i}^{-1}\operatorname{Tr}_{b}\Bigl[e^{+\beta\,\hat{H}_{s}}e^{-\beta\,\hat{H}_{c}}\,\bigl(\hat{H}_{s}+\hat{H}_{i}+\hat{H}_{b}\bigr)\Bigr]\,,\label{E:btsdssa}
\end{equation} 
then we obtain a new representation of $\mathcal{U}_{c}$
\begin{equation}
	\mathcal{U}_{c}=\operatorname{Tr}_{s}\Bigl\{\hat{\rho}_{s}\,\hat{U}_{i}\Bigr\}\,.
\end{equation}
Eq.~\eqref{E:btsdssa} is the quantum-mechanical version of \eqref{E:tbgd1} on account of the non-commutativity of the operators. In addition, we also note that $\hat{Z}_{i}$, an operator, is the quantum mechanical counterpart of $\mathcal{Z}_{i}$ in \eqref{E:rnrtr}. Since $\operatorname{Tr}_{s}\hat{\rho}_{s}=1$, we may alternatively define the operator $\hat{U}_{s}$ by
\begin{equation}\label{E:gkejsg}
	\hat{U}_{s}=\hat{U}_{i}-\mathcal{U}_{b}\,.
\end{equation}
such that
\begin{equation}\label{E:ngfddfs}
	\operatorname{Tr}_{s}\Bigl\{\hat{\rho}_{s}\,\hat{U}_{s}\Bigr\}=\operatorname{Tr}_{s}\Bigl\{\hat{\rho}_{s}\,\hat{U}_{i}\Bigr\}-\operatorname{Tr}_{s}\Bigl\{\hat{\rho}_{s}\,\mathcal{U}_{b}\Bigr\}=\mathcal{U}_{c}-\mathcal{U}_{b}=\mathcal{U}_{s}\,.
\end{equation}
Thus \eqref{E:rngkjfs} is recovered and the previous algebraic manipulations resemble their classical counterparts in Eqs.~\eqref{E:tbgd1}--\eqref{E:rtrfndeure}. The advantage of \eqref{E:btsdssa}, \eqref{E:gkejsg} is that, unlike \eqref{E:rtdnsq1}, \eqref{E:rtdnsqw}, they are all operators in the Hilbert space of the system $\mathbf{S}$. Indeed, using the identity operator $\hat{I}_{s}$ in the Hilbert space of the system $\mathbf{S}$ we can also define $\hat{U}_{b}$ as $\hat{U}_{b}=\mathcal{U}_{b}\,\hat{I}_{s}$.

In the same fashion, we may rewrite $\mathcal{A}_{c}$ in \eqref{E:ntrkjbs} by
\begin{align}
	\mathcal{A}_{c}=\operatorname{Tr}_{s}\biggl\{\frac{1}{Z}\operatorname{Tr}_{s}\Bigl[e^{-\beta\,\hat{H}}\,\hat{A}_{b}\Bigr]\biggr\}&=\operatorname{Tr}_{s}\biggl\{\hat{\rho}_{s}\,\hat{Z}_{i}^{-1}\operatorname{Tr}_{b}\Bigl[e^{+\beta\,\hat{H}_{s}}e^{-\beta\,\hat{H}}\,\hat{A}_{b}\Bigr]\biggr\}\,.
\end{align}
Thus we can define
\begin{equation}
	\hat{A}_{i}=\hat{Z}_{i}^{-1}\operatorname{Tr}_{b}\Bigl[e^{+\beta\,\hat{H}_{s}}e^{-\beta\,\hat{H}}\,\hat{A}_{b}\Bigr]\,,
\end{equation}
in an analogous form as \eqref{E:rrbgssdf1}, so that 
\begin{equation}
	\mathcal{A}_{c}=\operatorname{Tr}_{s}\Bigl\{\hat{\rho}_{s}\,\hat{A}_{i}\Bigr\}\,.
\end{equation}
We then can have a local form for the $\hat{A}_{s}$ given by
\begin{equation}\label{E:rnhjraz}
	\hat{A}_{s}=\hat{A}_{i}-\hat{A}_{b}\,,
\end{equation}
in close resemblance to \eqref{E:gnkjes}, if we re-define $\hat{A}_{b}$ as
\begin{equation}
	\hat{A}_{b}=\mathcal{A}_{b}\,\hat{I}_{s}\,.
\end{equation}
The expectation value of $\hat{A}_{s}$ is then
\begin{equation}\label{E:ruthggswq}
	\operatorname{Tr}_{s}\Bigl\{\hat{\rho}_{s}\,\hat{A}_{s}\Bigr\}=\operatorname{Tr}_{s}\Bigl\{\hat{\rho}_{s}\,\hat{A}_{i}\Bigr\}-\mathcal{A}_{b}=\mathcal{A}_{c}-\mathcal{A}_{b}=\mathcal{A}_{s}\,.
\end{equation}

Now we proceed with constructing a local form of the enthalpy operator of the system. From \eqref{E:gkejsg} and \eqref{E:rnhjraz}, we claim that the local form $\hat{\mathfrak{H}}_{s}$ is 
\begin{equation}
	\hat{\mathfrak{H}}_{s}=\hat{U}_{s}+J\cdot\hat{A}_{s}\,.
\end{equation}
We can straightforwardly show that
\begin{align}
	\operatorname{Tr}_{s}\Bigl\{\hat{\rho}_{s}\,\hat{\mathfrak{H}}_{s}\Bigr\}&=\operatorname{Tr}_{s}\Bigl\{\hat{\rho}_{s}\,\hat{U}_{s}\Bigr\}+J\cdot\operatorname{Tr}_{s}\Bigl\{\hat{\rho}_{s}\,\hat{A}_{s}\Bigr\}\notag\\
	&=\mathcal{U}_{s}+J\cdot\mathcal{A}_{s}
\end{align}
from \eqref{E:ngfddfs} and \eqref{E:ruthggswq}.

Thus we have succeeded in writing the operators that correspond to $\mathcal{A}_{s}$, $\mathcal{U}_{s}$, $\mathcal{H}_{s}$ in forms local in the Hilbert space of the system $\mathbf{S}$. {However, as can be seen from their expressions, their meanings are not transparent a priori. They are determined a posteriori because we would like their expectation values to take certain forms. This can pose a question about the uniqueness of these operators. At least for a given reduced density matrix $\hat{\rho}_{s}$ of the system, we can always attach an system operator $\hat{\Lambda}_{s}$ that satisfies $\operatorname{Tr}_{s}\hat{\rho}_{s}\hat{\Lambda}_{s}=0$ to the definitions of those local operators, that is, any system operator that has a zero mean}\footnote{The choice of $\hat{\Lambda}_{s}$ is not unique in the sense that in the basis $\{\lvert n\rangle\}$ that diagonalizes $\hat{\rho}_{\textsc{s}}$, we can write $\operatorname{Tr}_{s}\{\hat{\rho}_{\textsc{s}}\hat{\Lambda}_{\textsc{s}}\}=0$ as
\begin{equation}
	\operatorname{Tr}_{\textsc{s}}\{\hat{\rho}_{\textsc{s}}\hat{\Lambda}_{\textsc{s}}\}=\sum_{m,n}\langle n\vert\hat{\rho}_{\textsc{s}}\vert m\rangle\langle m\vert\hat{\Lambda}_{\textsc{s}}\vert n\rangle=\sum_{n}\bigl(\hat{\rho}_{\textsc{s}}\bigr)_{nn}\bigl(\hat{\Lambda}_{\textsc{s}}\bigr)_{nn}=0\,.
\end{equation}
It says that the vectors that are respectively composed of the diagonal elements of $\hat{\rho}_{\textsc{s}}$ and $\hat{\Lambda}_{\textsc{s}}$ are orthogonal, but it does not place any restriction on the off-diagonal elements of $\hat{\Lambda}_{\textsc{s}}$ in this basis.} .

So far, we essentially write the thermodynamic quantities by the quantum expectation value and in terms of the partition functions. Thus it is appropriate to introduce the Gibbs free energies of the composite $\mathbf{C}$, the system $\mathbf{S}$, and the bath $\mathbf{B}$, respectively by
\begin{align}
	\mathcal{G}_{c}&=-\beta^{-1}\ln\mathcal{Z}_{c}\,,&\mathcal{G}_{s}&=-\beta^{-1}\ln\mathcal{Z}_{s}\,,&\mathcal{G}_{b}&=-\beta^{-1}\ln\mathcal{Z}_{b}\,,
\end{align}
which obey the additive property of the Gibbs energy, $\mathcal{G}_{c}=\mathcal{G}_{s}+\mathcal{G}_{b}$. Futhermore, in the composite, we note that
\begin{align}
	\beta\bigl(\mathcal{H}_{c}-\mathcal{G}_{c}\bigr)&=\ln\mathcal{Z}_{c}-\beta\,\frac{\partial}{\partial\beta}\,\ln\mathcal{Z}_{c}=\beta^{2}\frac{\partial}{\partial\beta}\Bigl[-\beta^{-1}\,\ln \mathcal{Z}_{c}\Bigr]=\beta^{2}\frac{\partial \mathcal{G}_{c}}{\partial\beta}\,,\label{E:bgbjs1}
\end{align}
Meanwhile it can also be written as
\begin{align}
	\beta\bigl(\mathcal{H}_{c}-\mathcal{G}_{c}\bigr)=\ln \mathcal{Z}_{c}\,\operatorname{Tr}_{sb}\hat{\rho}-\frac{\beta}{\mathcal{Z}_{c}}\frac{\partial}{\partial\beta}\,\operatorname{Tr}_{sb}\Bigl\{e^{-\beta\,\hat{H}_{c}}\Bigr\}&=-\operatorname{Tr}_{sb}\Bigl\{\hat{\rho}_{c}\,\ln\hat{\rho}_{c}\Bigr\}\,.\label{E:bgbjs2}
\end{align}
From \eqref{E:bgbjs1} and \eqref{E:bgbjs2}, we can consistently define the entropy $\mathcal{S}$ of the composite by
\begin{equation}
	\mathcal{S}_{c}=\beta\bigl(\mathcal{H}_{c}-\mathcal{G}_{c}\bigr)=\beta^{2}\,\frac{\partial \mathcal{G}_{c}}{\partial\beta}=-\operatorname{Tr}_{sb}\Bigl\{\hat{\rho}_{c}\,\ln\hat{\rho}_{c}\Bigr\}\,.\label{E:nrtritu1}
\end{equation}
and, similarly, the entropy $\mathcal{S}_{b}$ of the bath:
\begin{equation}
	\mathcal{S}_{b}=\beta\bigl(\mathcal{H}_{b}-\mathcal{G}_{b}\bigr)=\beta^{2}\,\frac{\partial \mathcal{G}_{b}}{\partial\beta}=-\operatorname{Tr}_{b}\Bigl\{\hat{\rho}_{b}\,\ln\hat{\rho}_{b}\Bigr\}\,.\label{E:nrtritu2}
\end{equation}
From Eqs.~\eqref{E:nrtritu1} and \eqref{E:nrtritu2} the entropy $\mathcal{S}_{s}$ of the system in this representation is given by
\begin{equation}
	\mathcal{S}_{s}=\mathcal{S}_{c}-\mathcal{S}_{b}=\beta\bigl(\mathcal{H}_{s}-\mathcal{G}_{s}\bigr)=\beta^{2}\,\frac{\partial \mathcal{G}_{s}}{\partial\beta}=-\operatorname{Tr}_{sb}\Bigl\{\hat{\rho}_{c}\,\ln\hat{\rho}_{c}\Bigr\}+\operatorname{Tr}_{b}\Bigl\{\hat{\rho}_{b}\,\ln\hat{\rho}_{b}\Bigr\}\,,
\end{equation}
Note it is not equal to the von Neumann entropy, which is defined as the entropy of the system in the `bare' representation. 
\begin{equation}
	\mathcal{S}_{s}=-\operatorname{Tr}_{sb}\Bigl\{\hat{\rho}_{c}\,\ln\hat{\rho}_{c}\Bigr\}+\operatorname{Tr}_{b}\Bigl\{\hat{\rho}_{b}\,\ln\hat{\rho}_{b}\Bigr\}\neq-\operatorname{Tr}_{s}\Bigl\{\hat{\rho}_{s}\,\ln\hat{\rho}_{s}\Bigr\}\,.
\end{equation}

\section{Conclusion}\label{S:tihbgd}

\paragraph{Summary}
In this paper we provide quantum formulations for three systematized theories of thermodynamics at strong coupling, that {are proposed} by Gelin \& Thoss \cite{GT09}, Seifert \cite{Se16} and Jarzynski \cite{JA17}, respectively. All three formulations assume that the combined system + environment, which we call the composite, is initially in a global thermal state, that it remains in equilibrium for all times and is thus stationary. In such a configuration, even though the interaction between the system and the bath is non-negligible, the partition function of the composite is well defined. This facilitates the introduction of thermodynamic potentials in a way similar to the traditional vanishing-coupling thermodynamics.

In Approach I, {formulated} by Gelin and Thoss \cite{GT09}, one uses the intuitive definitions of the internal energy and the entropy, from which the thermodynamic relation among various thermodynamic potentials are established, but at the cost that these relations are much more complicated than their counterparts in the traditional vanishing-coupling thermodynamics. G\&T introduce an operator $\hat{\Delta}_{s}$ which signifies the deviation of the analytical forms of the thus-introduced thermodynamic potentials or relations from their conventional expressions as a consequence of strong system-bath coupling.  In contrast, Approach I\!I, adopted by Seifert \cite{Se16} for his classical formulation of thermodynamics at strong coupling, anchors on an attempt to preserve the thermodynamic relations between the thermodynamic potentials. The thermodynamic quantities of the system in this approach is defined by the differences between those of the composite  and those of the free bath. Thus the analytical expressions of these thermodynamic quantities, though still additive, are distinct from those in Approach I, and their physical interpretations become less transparent, in particular, in the quantum mechanics context.  

We further presented a quantum formulation of thermodynamics for classical systems strongly interacting with a bath when a fixed external agent is involved, as exemplified by Jarzynski's work \cite{JA17}. Such a configuration allows for the introduction of enthalpy, which may account for the work done by this fixed external agent switching on the coupling between the system and the bath. (This in fact has implicitly instilled the spirit of nonequilibrium processes  in the arguments that lead to the introduction of the enthalpy.) The effect of the bath operator linked to the external agent can be represented by an equivalent effect on the system, which then appears in the expression of the enthalpy of the system. Depending on how the system counterpart of the bath operator linked to the external agent is introduced, there are two representations, called the `bare' and `partial molar' representations by Jarzynski, corresponding respectively to the G\&T and Seifert approaches. We have worked out a quantum formulation for each of these two representations of Jarzynski's classical thermodynamics. 
 
\paragraph{Issues} We mention two outstanding issues of these two representations or approaches. When the quantum versions of these two approaches are applied to a small quantum system that strongly couples with a low-temperature bath, some nonintuitive results have been reported in the literature~\cite{NA02,HI06,HI08,OC06,HB08,HA11,IH09,HC17}. In Approach I, the von Neumann entropy is adopted as the system entropy, so when the system and the bath are entangled, this entropy will not approach zero for a simple system such as a harmonic oscillator in the zero temperature bath, contradicting the result in~\cite{HZ95}, where it has been shown that the ground state of such a composite system is non-degenerate in general, thus imply vanishing entropy at zero temperature. On the other hand, the heat capacity defined in Approach I\!I can take on negative values in the low temperature regime when the system consists of free particles or coupled harmonic oscillators. This anomalous behavior, not seen when the internal energy defined in Approach I is used, may be traced to an excessive inclusion of the interaction and the bath contributions, as shown in \eqref{E:rthrjf}, in the definition of the internal energy of the system.

\paragraph{Further Developments} 

We mentioned in the Introduction two major paradigms in quantum thermodynamics, closed system in a global thermal state (CGTs) approach, exemplified by all the systematics discussed in this paper, and the open-system nonequilibrium (ONEq) approach, such as in \cite{HC17} (see references therein).  Each has its special merits and limitations. If one can have the quantum formulations of the CGTs and the ONEq approaches in place,  one can proceed to explore the thermodynamic laws involving quantum energy and entropy, while taking advantage of the ease in defining work for a closed system in the globally thermal state set-up. One can also take advantage of the ONEq approach in seeing how a quantum system strongly interacting with a quantum bath evolves explicitly in time. We point out a few places where development toward this goal is made possible from our present investigation: 

1) In the two separate approaches (I \& I\!I), we have identified \textit{the expressions for the reduced density matrix for the system}. This will enable us to compare the physical quantities of interest, such as the nonequilibrium evolution of the system dynamics between these two paradigms.

2) It would be interesting to see if some thermodynamic function like the free energy may exist in the nonequilibrium setting  even though the partition function and the thermodynamic quantities defined therefrom are not.  This may enable one to \textit{define free energy density via the generating functional ab initio}, up-lifting this very important and useful thermodynamic function from the restricted weak-coupling equilibrium thermodynamics to the fully nonequilibrium, strong coupling conditions.  An observation from~\cite{SFTH,HC17} encourages us to pursue this inquiry. There it was shown that for a quantum system started from a nonequilibrium initial state and bilinearly coupled to a thermal quantum bath, after relaxation in a nonequilibrium evolution (i.e., in the ONEq set up) its reduced density matrix will approach the reduced density matrix of the system derived from assuming that the closed system + bath stay in a global thermal state (i.e., in the CGTs set-up). This indicates that the generating functional of the reduced system in the nonequilibrium open system will be the partition function $\mathcal{Z}^*$ after the reduced system is relaxed to an equilibrium state. It also suggests that the free energy $\mathcal{F}^*$ can the free energy we are looking for. We will explore this line of reasoning in the influence functional framework used recently in \cite{HC17} in the open-systems nonequilibrium dynamics treatment of strong coupling quantum thermodynamics.

3) A possible development of the operator  $\hat{\Delta}_{s}$ introduced by Gelin and Thoss, and the Hamiltonian operator of mean force discussed in Approach I\!I, can be related to the \textit{influence action or the coarse-grained effective action} of the system when they are sandwiched by the states of the system and formulated in the imaginary-time path integral method. This and the earlier observation we made for the partition function provide us with sufficient motivation to extend the present CGTs equilibrium formulation to a nonequilibrium framework by employing the real-time closed-time-path formalism used in ~\cite{HC17} and by others. We hope to report these results in our next paper in this series \cite{HH18}.

\vskip .25cm

\noindent {\bf Acknowledgment}  We thank Prof C. Jarzynski for explaining the key points in his work, and Prof. C. H. Chou  and Dr. Y. Suba\c{s}\i{} for helpful discussions. BLH visited the Fudan University Physics Theory Center when this work commenced, while JTH visited the University of Maryland Physics Theory Center when this work was concluded.

\vskip .5 cm

\appendix
\section{Handling Operator Products in Quantum Thermodynamics}\label{S:opera}

In deriving various thermodynamic relations in the context of QTD, we ofter come up with expression involving exponential of the sum of two operators, say $\hat{\lambda}$ and $\hat{\mu}$. Unlike its $c$-number counterpart, in general such an exponential cannot be written as a product of exponential of the respective operators, that is,
\begin{equation}\label{E:trutrer}
	e^{\hat{\lambda}+\hat{\mu}}\stackrel{?}{=} e^{\hat{\lambda}}\, e^{\hat{\mu}}\,,
\end{equation}
because these two operators $\hat{\lambda}$, $\hat{\mu}$ may not commute. From Baker-Campbell-Haussdorff formulas, the righthand side of \eqref{E:trutrer} in fact is given by
\begin{equation}\label{E:erhgggfdbjgs}
		e^{\hat{\lambda}}e^{\hat{\mu}}=\exp\biggl(\hat{\lambda}+\hat{\mu}+\frac{1}{2!}\,\bigl[\hat{\lambda},\hat{\mu}\bigr]+\frac{1}{3!}\,\Bigl\{\frac{1}{2}\,\bigl[\bigl[\hat{\lambda},\hat{\mu}\bigr],\hat{\mu}\bigr]+\frac{1}{2}\,\bigl[\hat{\lambda},\bigl[\hat{\lambda},\hat{\mu}\bigr]\bigr]\Bigr\}+\cdots\biggr)\,,
\end{equation}
for any two operators $\hat{\lambda}$, $\hat{\mu}$. However, in the special case that $[\hat{\lambda},\hat{\mu}]=0$, the equality in \eqref{E:trutrer} indeed is valid. The other useful expression in the Baker-Campbell-Hausdorff formulas is
\begin{equation}\label{E:erhfdbjgs}
		e^{\hat{\lambda}}\,\hat{\mu}\,e^{-\hat{\lambda}}=\hat{\mu}+\bigl[\hat{\lambda},\hat{\mu}\bigr]+\frac{1}{2}\,\bigl[\hat{\lambda},\bigl[\hat{\lambda},\hat{\mu}\bigr]\bigr]+\cdots\,.
\end{equation}
This is particular useful in deriving the unitary transformation of $\hat{\mu}$ by the  unitary operator $e^{\hat{\lambda}}$.

We also often come to a situation that we need to take a derivative of an exponential of the operator. This is less straightforward than is expected due to the fact that the operator in the exponent may not commute with its own derivative. For example, consider an operator $\hat{O}(\chi)$ of the form
\begin{equation}
	\hat{O}(\chi)=\alpha(\chi)\,\hat{X}+\beta(\chi)\,\hat{P}\,,
\end{equation}
where $\alpha$, $\beta$ are functions of $\chi$, but the operators $\hat{X}$, $\hat{P}$ of the canonical variables have no explicit $\chi$ dependence. We immediately see the trivial result $[\hat{O}(\chi),\,\hat{O}(\chi)]=0$, but
\begin{equation}
	\bigl[\hat{O}(\chi),\partial_{\chi}\hat{O}(\chi)\bigr]=\bigl(\alpha\dot{\beta}-\dot{\alpha}\beta\bigr)\,\bigl[\hat{X},\hat{P}\bigr]\neq0\,,
\end{equation}
where the overhead dot represents the derivative with respect to $\chi$. This introduces complications in taking the derivative of, say, $e^{-\hat{O}(\chi)}$ with respect to $\chi$. If we realize an operator function in terms of its Taylor's expansion, then
\begin{equation}
	e^{-\hat{O}(\chi)}=\sum_{k=0}^{\infty}\frac{(-1)^{k}}{k!}\,\hat{O}^{k}(\chi)\,.
\end{equation}
Taking the derivative with respective to $\chi$, we have the righthand side given by
\begin{align}
	&\quad\partial_{\chi}e^{-\hat{O}}\notag\\
	&=-\sum_{k=1}^{\infty}\frac{(-1)^{k-1}}{(k-1)!}\,\biggl\{\frac{1}{k}\,\Bigl[\bigl(\partial_{\chi}\hat{O}\bigr)\underbrace{\hat{O}\cdots\cdots \hat{O}}_{(k-1)\, \text{terms}}+\hat{O}\bigl(\partial_{\chi}\hat{O}\bigr)\underbrace{\hat{O}\cdots\cdots \hat{O}}_{(k-2)\, \text{terms}}+\cdots\cdots+\underbrace{\hat{O}\cdots\cdots \hat{O}}_{(k-1)\, \text{terms}}\bigl(\partial_{\chi}\hat{O}\bigr)\Bigr]\biggr\}\notag\\
	&=-\sum_{k=1}^{\infty}\frac{(-1)^{k-1}}{(k-1)!}\,\Bigl[\bigl(\partial_{\chi}\hat{O}\bigr)\hat{O}^{k-1}\Bigr]_{\text{sym}}\notag\\
	&=-\Bigl[\bigl(\partial_{\lambda}\hat{O}\bigr)\,e^{-\hat{O}}\Bigr]_{\text{sym}}\,,\label{E:trbdfg}
\end{align}
where we define the symmetrized product $(\hat{O}_{1}\hat{O}_{2}\cdots \hat{O}_{k})_{\text{sym}}$ as a generalization of the anti-commutator by
\begin{equation}
	(\hat{O}_{1}\hat{O}_{2}\cdots \hat{O}_{k})_{\text{sym}}=\frac{1}{\text{\# of perm.}}\sum_{\text{\# of perm.}}\hat{O}_{\sigma_{1}}\hat{O}_{\sigma_{2}}\cdots \hat{O}_{\sigma_{k}}\,,
\end{equation}
with $\sigma$ being the permutations of $1$, $2$, $\cdots$, $k$. Thus the expression $(\partial_{\beta}\hat{H}_{s}^{*})\,e^{-\beta\,\hat{H}_{s}^{*}}$ in \eqref{E:dfkekrfd} and similar expressions in the subsequent paragraphs will be understood in this manner as a symmetrized product of $\partial_{\beta}\hat{H}_{s}^{*}$ and the Taylor-expanded $e^{-\beta\,\hat{H}_{s}^{*}}$, shown in \eqref{E:trbdfg}.

However if the derivative like \eqref{E:trbdfg} is taken within a trace, then the complicated expression \eqref{E:trbdfg} will reduce to a simple form
\begin{align}
	&\quad\operatorname{Tr}\bigl\{\partial_{\chi}e^{-\hat{O}}\bigr\}\notag\\
	&=-\sum_{k=1}^{\infty}\frac{(-1)^{k-1}}{(k-1)!}\,\biggl[\frac{1}{k}\,\operatorname{Tr}\Bigl\{\bigl(\partial_{\chi}\hat{O}\bigr)\underbrace{\hat{O}\cdots\cdots\hat{O}}_{(k-1)\, \text{terms}}+\hat{O}\bigl(\partial_{\chi}O\bigr)\underbrace{\hat{O}\cdots\cdots\hat{O}}_{(k-2)\, \text{terms}}+\cdots\cdots+\underbrace{\hat{O}\cdots\cdots\hat{O}}_{(k-1)\, \text{terms}}\bigl(\partial_{\chi}\hat{O}\bigr)\Bigr\}\biggr]\notag\\
	&=-\sum_{k=1}^{\infty}\frac{(-1)^{k-1}}{(k-1)!}\,\operatorname{Tr}\Bigl\{\bigl(\partial_{\chi}\hat{O}\bigr)\,\hat{O}^{k-1}\Bigr\}\notag\\
	&=-\operatorname{Tr}\Bigl\{\bigl(\partial_{\chi}\hat{O}\bigr)\,e^{-\hat{O}}\Bigr\}\,,\label{E:tbgfbsd}
\end{align}
due to the cyclic property of the trace formula. Hence in general we have
\begin{equation}
	\partial_{\chi}e^{-\hat{O}}=-\Bigl[\bigl(\partial_{\lambda}\hat{O}\bigr)\,e^{-\hat{O}}\Bigr]_{\text{sym}}\,,
\end{equation}
but the trace of it
\begin{equation}\label{E:ritjirwe}
	\operatorname{Tr}\bigl\{\partial_{\chi}e^{-\hat{O}}\bigr\}=-\operatorname{Tr}\bigl\{\Bigl[\bigl(\partial_{\lambda}\hat{O}\bigr)\,e^{-\hat{O}}\Bigr]_{\text{sym}}\bigr\}=-\operatorname{Tr}\bigl\{\bigl(\partial_{\chi}\hat{O}\bigr)\,e^{-\hat{O}}\bigr\}\,,
\end{equation}
as if the operator $\hat{O}$ is a $c$-number. Note here we have assumed the traces applied in \eqref{E:tbgfbsd}--\eqref{E:ritjirwe} are not a partial trace; otherwise the same symmetrization procedure is still necessary.

A special case of \eqref{E:tbgfbsd} is 
\begin{equation}
	\partial_{\chi}e^{-\chi\,\hat{O}}
\end{equation}
where $\hat{O}$ has no explicit dependence on $\chi$. Then it is straightforward to perform the differentiation, and we obtain
\begin{equation}\label{E:rtbrjdf}
	\partial_{\chi}e^{-\chi\,\hat{O}}=-\hat{O}\,e^{-\chi\,\hat{O}}\,,
\end{equation}
since $[\hat{O},e^{-\chi\,\hat{O}}]=0$.

Next we give an explicit application of \eqref{E:tbgfbsd} to the derivation of \eqref{E:lpernere}. In particular, we focus on the expression
\begin{align}
	&\quad-\frac{1}{\mathcal{Z}_{c}}\frac{\partial}{\partial\beta}\operatorname{Tr}_{s}e^{-\beta(\hat{H}_{s}+\hat{\Delta}_{s})}\,,&&\text{with}&\mathcal{Z}_{c}&=\operatorname{Tr}_{s}e^{-\beta(\hat{H}_{s}+\hat{\Delta}_{s})}\,.\label{E:prrnfg}
\intertext{Carrying out the differentiation of \eqref{E:prrnfg} gives}
	&=\frac{1}{\mathcal{Z}_{c}}\operatorname{Tr}_{s}\Bigl\{\bigl(\hat{H}_{s}+\hat{\Delta}_{s}+\beta\,\partial_{\beta}\hat{\Delta}_{s}\bigr)e^{-\beta(\hat{H}_{s}+\hat{\Delta}_{s})}\Bigr\}\label{E:trijrngfs}\\
	&=\langle \hat{H}_{s}\rangle+\langle\hat{\Delta}_{s}\rangle+\beta\,\langle\partial_{\beta}\hat{\Delta}_{s}\rangle\,.
\end{align}
The first two terms in \eqref{E:trijrngfs} is the consequence of \eqref{E:rtbrjdf}, while the third term results from \eqref{E:ritjirwe} due to the trace.

\end{document}